\documentclass[11pt,twoside]{article}
\usepackage{amssymb}
\usepackage[mathscr]{eucal}
\usepackage{eufrak}
\usepackage{amsmath,amsthm}
\usepackage{mathrsfs}
\usepackage{lgreek}
\usepackage[shortlabels]{enumitem}
\usepackage[colorlinks=true,
   urlcolor=blue,           filecolor=green,      
   citecolor=green,      
   linkcolor=red,           bookmarks=true,
  unicode,
   plainpages=false,   ]{hyperref}
\usepackage{color}

\definecolor{royalblue}{rgb}{0,0,0.128}
\def\bp{\begin{proof}}
\def\ep{\end{proof}}
\def\n{\nabla}
\def\ssfrac#1#2{\mbox{\large$\frac{#1}{#2}$}}
\def\sfrac#1#2{\mbox{\Large$\frac{#1}{#2}$}}
\def\intl#1{\int\limits_{#1}}
\def\intll#1#2{\int\limits_{#1}^{#2}}
\def\dm{|\hskip-0.05cm|}

\def\OO{\Omega}
\def\displ{\displaystyle}
\def\VSE{\vspace{6pt}\\&\displ }
\def\VS{\vspace{6pt}\\\displ }
\def\rf#1{{\rm(\ref{#1})}}

\def\R{\Bbb R}
\def\N{\Bbb N}
\def\à{à}

\def\vep{\varepsilon}

\def\be{\begin{equation}}
\def\ba{\begin{array}}
\def\ea{\end{array}}
\def\ee{\end{equation}}
\def\vs1{\vspace{1ex}}

\def\ov{\overline}

\def\po{\partial\Omega}

\def\Ã©{\'{e}}
\def\Ãš{\`{e}}
\newtheorem{lemma}
{\bf Lemma} 
\font\sc=cmcsc10
\title{\Large\bf On the  weak solutions to the Navier-Stokes equations:\\a possible gap related to the energy equality}
\author{\sc  Paolo Maremonti
\thanks{Dipartimento di Matematica e Fisica,  
Universit\`{a} degli 
Studi della Campania
``L. Vanvitelli'', via Vivaldi 43, 81100 \null\hskip0.55cmCaserta,
 Italy.
paolo.maremonti@unicampania.it\newline\null\hskip0.55cm The  research activity  is performed under the
auspices of   GNFM-INdAM. }}
\date{\it to Senjo Shimizu on her 60th birthday}
\begin{document}
\markboth{\footnotesize\rm    P. Maremonti} {\footnotesize\rm
On the  weak solutions to the Navier-Stokes equations: a possible 
gap related to the energy equality}
\maketitle 
{\small\bf Abstract} - {\small It is well known that a Leray-Hopf weak solution enjoys  an energy inequality. Here,  we investigate  the energy equality   related to a suitable  weak solution to the Navier-Stokes initial boundary value problem.  The term suitable is meant  in the sense that for our goals we achieve a weak solution whose existence is based as limit of solutions to the mollified Navier-Stokes system.  In the case of a weak regularity of the solution,  our results  justify the possible gap  for the energy equality in  terms  of ``kinetic energy''. However, 
if there is a sufficient regularity, {\it e.g.}, like the continuity of the $L^2$-norm  of the weak solution, then the energy equality holds. }
\noindent
\newcommand{\red}{\protect\bf}
\renewcommand\refname{\centerline
{\red {\normalsize \bf References}}}
\newtheorem{ass}
{\bf Assumption} 
\newtheorem{defi}
{\bf Definition} 
\newtheorem{tho}
{\bf Theorem} 
\newtheorem{rem}
{\sc Remark} 
\newtheorem{coro}
{\bf Corollary} 
\newtheorem{prop}
{\bf Proposition} 
\renewcommand{\theequation}{\arabic{equation}}
\setcounter{section}{0}
\section{Introduction}
We consider the  Navier-Stokes initial boundary value problem:
\be\label{NS}\ba{cc}v_t+v\cdot\n v+\n\pi=\Delta v\,,&\n\cdot v=0\,,\mbox{ in }(0,T)\times \OO\,,\VS v=0\mbox{ on }(0,T)\times\po\,,& v=v_0\mbox{ on }\{0\}\times\OO\,,\ea\ee where $\OO\subseteq\R^3$ can be a   bounded or an exterior  domain, whose boundary $\po$ for simplicity 
is assumed smooth, a half-space or the whole space. We set $w_t:=\frac\partial{\partial t}w$ and $w\cdot\n w:= (w\cdot\n) w$. \par We investigate   the existence of a weak solution $v$, in the sense of Leray-Hopf, to problem \rf{NS} enjoying the energy equality in the following form:
\be\label{EE}\dm v(t)\dm_2^2+2\intll 0t\dm \n v(\tau)\dm_2^2d\tau=\dm v_0\dm_2^2\,,\;t >0\,.\ee
In 2D  the result is true for all $t>0$. Instead,  the validity of the \rf{EE} is an open problem in $n$D, $n\geq3$. Actually, limiting ourselves to the three-dimensional case,  a weak solution {\it a priori} satisfies an energy inequality. The following one is in the strong form\footnote{\,The energy inequality \rf{NS-III} is said in strong form in contraposition to the one satisfied {\it a priori} only for $s=0$, called in weak form. This is the case $n$D of a Hopf's weak solution to the IBVP in unbounded domains, where no energy   inequality   is  known in the strong form.}   and is due to Leray (1934) \cite{L}:
\be\label{NS-III}\dm v(t)\dm_2^2+2\intll st\dm \n v(\tau)\dm_2^2d\tau\leq \dm v(s)\dm_2^2\,,\mbox{ for all }t>s\mbox{ and a.e. in } s>0 \mbox{ and for }s=0\,.\ee 
{\it A priori} there is no reason for the validity of a  strong inequality in \rf{NS-III}. To date, it appears as a consequence of the weakness of the weak solution $v$. \par  In  Leary's  paper \cite{L}, the possible instants of singularity  are heuristically interpreted as possible phenomenas of turbulence in a fluid motion, that are expected in  3D.\par We know, in part also thanks to inequality \rf{NS-III},   that the set $S$ of the possible instants of singularity of a Leray weak solution has at least $\mathcal H^\frac12(S)=0$, $\mathcal H^a$ is the $a$-dimensional Hausdorff measure, \cite{CKN,Sh}.
\par In order to state our results and to attempt a comparison with the ones on the  topic  already known in literature, we  introduce some notation and the definition of weak solution.
\par By the symbol $\mathscr C_0(\OO)$ we mean the set of functions divergence free and belonging to $C_0^\infty(\OO)$. We indicate by $J^2(\OO)$ and by $J^{1,2}(\OO)$ the completion of $\mathscr C_0(\OO)$  in $L^2(\OO)$ and in $W^{1,2}(\OO)$, respectively. \par We denote by $|D|$ the finite measure of a Lebesgue  measurable set $D$. Moreover, for any family $\mathscr F:= \{X_i:X_i\subset \R \mbox{ with } i\in\mathscr I\subseteq \N\}$ by $|\mathscr F|$ we mean the Lebesgue measure of the set $\underset {i\in \mathscr I}\cup X_i$.\par We use the symbol  $\rightharpoonup$ and the symbol $\to$ to mean the weak and the strong convergence, respectively.\begin{defi}\label{DWS}{\sl For all $v_0\in J^2(\OO)$, a field $v:(0,\infty)\times\OO\to\R^3$ is said a weak solution corresponding to  datum $v_0$ if \begin{itemize}\item[{\rm1)}]for all $T>0$, $v\in L^\infty(0,T;J^2(\OO))\cap L^2(0,T;J^{1,2}(\OO))$,\item[{\rm 2)}]for all $T>0$ and  $t,s\in(0,T)$,  the field $v$
satisfies the integral   equation:
\newline \centerline{$\displ\intll
st\Big[(v,\varphi_\tau)-(\nabla
v,\nabla
\varphi)+(v\cdot\nabla\varphi,v)\Big]d\tau+(v(s),\varphi
(s))=(v(t),\varphi(t))$,} for all $\varphi(t,x)\in C_0^1([0,T)\times\OO)$ with $\n\cdot\varphi(t,x)=0$\,,\item[{\rm 3)}]$\displ\lim_{t\to0}\,(v(t),\varphi)=(v_0,\varphi)\,,\mbox{ for all }\varphi\in\mathscr C_0(\OO)\,.$\end{itemize}}\end{defi} The following theorem is well known in literature, and for the Cauchy problem it was proved by Leray in \cite{L}.  \begin{tho}\label{CTII}{\sl For all $v_0\in J^2(\OO)$ there exists a weak solution $v$ to problem \rf{NS} that enjoys the properties: 
$$\ba{c}\displ\dm v(t)\dm_2^2+2\!\intll st\!\dm \n v(\tau)\dm_2^2d\tau\leq \dm v(s)\dm_2^2\,,\\\mbox{\,for\,all\,}t>s,\mbox{\,and\,for\,all\,}s\!\in\! [0,\infty)-I,\,I\hskip-0.04cm\subset\!(0,\theta), \,\theta\leq c\dm v_0\dm_2^4\,, \,|I|=0\,,\,c\mbox{\,independent\,of\,}v_0\,;\vspace{3pt}\ea$$ \be\label{SLO} \mbox{for\,all\,}s\!\notin\! I,\lim_{t\to s^+}\dm v(t)-v(s)\dm_2=0\,;\,\ee $$\mbox{for\,all\,}\varphi\!\in\! J^2(\OO)\,,	\; (v(t),\varphi)\mbox{\,is\,a\,continuous\,function\,of }t\,.
$$   
}\end{tho}
\begin{defi}\label{LS}{\sl A solution of Theorem\,\ref{CTII} is said a {\it Leray weak solution.}}\end{defi}
\begin{prop}\label{PPE}{\sl Let $v$ be a Leray weak solution. Then we get \be\label{PP-I}\dm v(t)\dm_2^2+2\intll 0t\dm \n v(\tau)\dm_2^2d\tau=\dm v_0\dm_2^2\,,\mbox{ a.e. in }t >0\,, \ee if, and only if, 
\be\label{PP-II}\dm v(t)\dm_2^2+2\intll st\dm \n v(\tau)\dm_2^2d\tau=\dm v(s)\dm_2^2\,,\;\mbox{a.e. in }t >s\,.\ee
}\end{prop} Although the proof of Proposition\,\ref{PPE} is immediate,   we develop one in sect.\,\ref{PPTC}.\par
In the following by the symbol $\lfloor\,\cdot\,\rfloor$ we mean the  integer part of a real number. \par
The following theorem is related to our chief results:
\begin{tho}\label{CTI}{\sl For all $v_0\in J^2(\OO)$ there exists a Leary weak solution $v$ that admits one and only one of the following two alternatives:  almost everywhere in $t>0$ the energy equality \rf{PP-I} holds, or,  almost everywhere in $t, s\geq0$,  there exists an one-parameter family of non-negative  integers $\{\mbox{\rm\texttt{n}}(\alpha)\}$ such that \begin{itemize}\item if ${\mbox{\rm\texttt{n}}}(\alpha)\in \N$  the following special energy equality holds: \vspace{-5pt}\be\label{CTII-II}\hskip-0.5cm\dm v(t)\dm_2^2+2\!\!\intll st\!\!\dm\n v(\tau)\dm_2^2d\tau+\!\!\lim_{\alpha\to1^-}\mbox{$\underset{h=1}{\overset{\mbox{\scriptsize\rm\texttt{n}}(\alpha)}\sum}$}\Big[\dm v(s_h(\alpha))\dm_2^2-\dm v(t_h(\alpha))\dm_2^2\Big]\!=\dm v(s)\dm_2^2\,,\ee  
where ${\mbox{\rm\texttt{n}}}(\alpha)\leq\lfloor 2c\hskip0.01cm\tan\alpha\frac\pi2\rfloor$, for all $\alpha\in(\alpha_0',1)$, $\alpha_0'>0$,          $c$ is a constant independent of $v_0$, $s,\,t$, and   $\{(s_h(\alpha),t_h(\alpha))\}_{h=1,\cdots,\mbox{\footnotesize\rm\texttt{n}}(\alpha)}=:\widetilde J(\alpha)\subset (s,t)$ is a family of pairwise disjoint    open intervals enjoying 
the limit properties $$\displ\lim_{\alpha\to1^-}(1-\alpha)^{-1}|\widetilde J(\alpha)|= 0\mbox{ \,and }\lim_{\alpha\to1^-}(1-\alpha)\hskip0.03cm\mbox{\rm\texttt n}(\alpha)=0\,,$$  \item if ${\mbox{\rm\texttt{n}}}(\alpha)=0$ for all $\alpha\in(\alpha'_0,1)$,\,$\alpha'_0>0$, then on $(s,t)$ the energy equality \rf{PP-II} holds.\end{itemize} }  
\end{tho}
\begin{coro}\label{CoroC}{\sl The energy equality \rf{PP-I} holds if, and only if, almost everywhere in $s$ and $t$ there exists a $\alpha'_0\in(0,1)$ such that ${\rm\texttt n}(\alpha)=0$ for all $\alpha\in(\alpha'_0,1)$.}\end{coro}
\begin{rem}\label{RI}{\rm \phantom\newline\begin{itemize}\item 
Our result is an existence result, so that we suitably construct our weak solution $v$ of Theorem\,\ref{CTII} and Theorem\,\ref{CTI}. The starting point is a sequence  $\{v^m\}$, whose $m$th element is a smooth  solution to the {\it mollified}  Navier-Stokes IBVP (see \rf{NSM}), this is the Leray approach to the existence of a weak solution.  In our proof we  prove that, for all $q\in[1,2)$, the sequence $\{\n v^m\}$   converges strong in $L^q(0,T;L^2(\OO))$. This result was proved for the first time in \cite{CGM-I} (see also Lemma\,\ref{SCLD} of this note) and it is crucial for our aims. \item In connection with a Leray's weak  solution    the following elucidation is due. If for Leary's weak solution we mean the one given in Definition\,\ref{LS}, whose set of solutions, by virtue of Theorem\,\ref{CTII}, is certainly not empty, then, being not known a uniqueness theorem, {\it a priori} we cannot say that a Leray's weak solution enjoys  the properties of Theorem\,\ref{CTI}. Instead, if for Leray's weak solution we mean the one that is constructed as limit of the sequence $\{v^m\}$, where $v^m$ is a solution to the mollified Navier-Stokes system \rf{NSM}, then, by virtue of the uniqueness of weak limit, then we can say that it enjoys the properties of Theorem\,\ref{CTI}.\item By virtue of Proposition\,\ref{PPE}, if, and only if, there exists an interval $(s,t)$ for which the integer $\mbox{\rm\texttt{n}}(\alpha)$ belongs to $\N$, for all $\alpha\in(\alpha'_0,1)$, then the energy equality fails  to hold. However, we are not able to detect a such interval.  \item It is natural to inquire about the claim   $t>s\geq0$ a.e. for the energy equalities 
\rf{CTII-II}, that does not exclude that for $s=0$ equality \rf{CTII-II} does not hold. The instants $s<t$ that ensure \rf{CTII-II}, indicated by $\mathcal T$ their set,   are the ones for which $\dm \n v^m(t)\dm _2$, related to the sequence $\{v^m\}$ of solutions to \rf{NSM}, admits a bound uniform with respect to $m$ (see \rf{GUB}). Being $\mathcal T\subseteq I^c$ ($I$ is stated in Theorem\,\ref{CTII}), the instants $s\in I^c-\mathcal T$ appear lacking.  However, recalling that our weak solution is continuous in $L^2$-norm on the right of  $s\in I^c$, one achieves the result for $s\in I^c$, in particular if $v_0\in J^2(\OO)-J^{1,2}(\OO)$, by means of a suitable sequence of instant $\{s_p\}\to s$ and by considering the limit on $p$ of the related terms in  
\rf{CTII-II}. 
\item The special bound $\lfloor 2c\tan{\alpha\frac\pi2}\rfloor$ for the integers $\texttt{n}(\alpha)$ is due to the special auxiliary function that we employ to achieve the result. However,     one can consider other auxiliary functions for the goal. In this regard it is worth to stress that they lead to the same result. In our proof, we employ the auxiliary  function that seems to be  simplest to us for the computations. \par However, we do not consider it  useful to state our result looking for a  function enjoying suitable qualitative  properties to achieve the result. Because here we want and we, coherently,  need to quantify the possible gap for the  energy equality. So that for us it is more interesting any particular function that is  better at quantify the gap. \item In Theorem\,\ref{CTI}  the energy equality holds if   $\mbox{\rm\texttt{n}}(\alpha)=0$ for all $\alpha\in(\alpha_0',1)$. However, \texttt{n}$(\alpha)\ne0$ is a possible integer. Actually, we can also achieve another integer:  $ \texttt{N}(\alpha)\in [\texttt{n}(\alpha), \lfloor 2c\tan\alpha\frac\pi2\rfloor]$. The integer \texttt{n}$(\alpha)$ is deduced as min-limit on $m$ of a suitable sequence $\{$\texttt{n}$  (\alpha,m)\}$,  Lemma\,\ref{CJAM}, instead \texttt{N}$(\alpha,m)$ is deduced as max-limit of the same sequence. Considering the subsequence that achieves the max-limit $\texttt{N}(\alpha)$ and developing the same argument lines employed for the subsequence related to the min-limit  \texttt{n}$(\alpha)$ (see Lemma\,\ref{NUP}), one formally obtains the same result of Theorem\,\ref{CTII}, that is 
$$\dm v(t)\dm_2^2+2\intll st\dm\n v(\tau)\dm_2^2d\tau+\lim_{\alpha\to1^-}\mbox{$\underset{h=1}{\overset{\mbox{\scriptsize\rm\texttt{N}}(\alpha)}\sum}$}\left[\dm v(\sigma_h(\alpha))\dm_2^2-\dm v(\tau_h(\alpha))\dm_2^2\right]=\dm v(s)\dm_2^2\,,$$
where, for all $\alpha\in(\alpha_0',1)$, $\{(\sigma_h(\alpha),\tau_h(\alpha))\}_{h=1,\cdots,\mbox{\footnotesize\rm\texttt{N}}(\alpha)}$ is a family of pairwise disjoint   open intervals. 
Finally,  set ${\widehat J}(\alpha):=\{(\sigma_h(\alpha),\tau_h(\alpha))\mbox{ for }h=1,\cdots,\mbox{\rm\texttt{N}}(\alpha)\}$,
one gets the limit properties  $$\lim_{\alpha\to1^-}(1-\alpha)^{-1}|\widehat J(\alpha)|=0\,\mbox{ \,and }\lim_{\alpha\to1^-}(1-\alpha)\hskip0.03cm\mbox{\rm\texttt N}(\alpha)=0\,.$$ We are not able to compare $\widehat J(\alpha)$ and $\widetilde J(\alpha)$ of Theorem\,\ref{CTI}.
\item
In Theorem\,\ref{CTI} the possible gap is given in terms of the {``}kinetic energy'' $\dm v(t)\dm_2^2$. Via \rf{CTII-II}, letting $\alpha\to1^-$, we have $|t_h(\alpha)-s_h(\alpha)|\to0$, then in the case of $\displ\lim_{\alpha\to1^-}\dm v(s_h(\alpha))\dm_2^2-\dm v(t_h(\alpha))\dm_2^2>0$  we read into a  discontinuity in time of the $L^2$-norm of the solution $v$.\item   For a Leray weak solution formula \rf{CTII-II} is a possible  gap for the energy equality. However, we can achieve another, that is the one given in   \rf{EE-N-I} that we point out in Remark\,\ref{FRE}. This formula furnishes the possible gap by means of the {``}internal energy''. \item If for some $s,\,t\in\cal T$ the energy equality \rf{PP-II} does not hold,  then for all the extracts  $\{\n v^{m_k}\}_{k\in\N}$ the strong convergence in $L^2(0,T;L^2(\OO))$ does not hold. Then one easily gets that there exists an extract such that 
$$\lim_{m_k}\intll st\dm \n v^{m_k}-\n v\dm_2^2d\tau=\lim_{\alpha\to1^-}\mbox{$\underset{h=1}{\overset{\mbox{\scriptsize\rm\texttt{n}}(\alpha)}\sum}$}\Big[\dm v(s_h(\alpha))\dm_2^2-\dm v(t_h(\alpha))\dm_2^2\Big]\,,$$or equivalently $$\lim_{m_k}\intll st\dm \n v^{m_k}\dm_2^2d\tau= \intll st\dm \n v\dm_2^2d\tau+\lim_{\alpha\to1^-}\mbox{$\underset{h=1}{\overset{\mbox{\scriptsize\rm\texttt{n}}(\alpha)}\sum}$}\Big[\dm v(s_h(\alpha))\dm_2^2-\dm v(t_h(\alpha))\dm_2^2\Big]\,.$$\end{itemize}}\end{rem}
   \begin{rem}\label{R-I}{\rm \phantom\newline\begin{itemize}
\item We set the viscosity  $\nu=1$, not being interested to the questions connected with the vanishing viscosity. Hence, in our results, for the constants we do not explicit   the dependence on the viscosity $\nu$ of the fluid. However, if $\nu\to0$ some results fail to hold.     \item We  stress that in Theorem\,\ref{CTI} the claim related to the validity of \rf{NS-III} has not to be considered pleonastic. Actually, assuming $s=0$, estimate \rf{NS-III} holds    for all $t>0$, in contrast to the validity {\it a priori} almost everywhere of the special energy equalities  \rf{CTII-II}. \item We think that having clarified in \rf{CTII-II} the possible gap in the   energy equality, it is now possible to better  delimit  the ambit of the question of an extra-dissipation. In particular, if   some motive of weakness  of the solution or of turbulence phenomena  of the fluid  makes happen    an ``extra energy'' that is not equivalent to the one exhibits in \rf{CTII-II}, then we have a case of  non-uniqueness for weak solutions.\item Related to the last remark, that is connection between energy relation and uniqueness question, we would like to point out the following. In \cite{BV} the authors furnish an example of  non uniqueness for very weak solutions $u$ that are continuous in $L^2$-norm. The example does not work provided that the very weak solution $u$ belongs to $L^2(0,T;J^{1,2}(\OO))$. Hence, the energy relation can be discriminant for the uniqueness question. This is the case for very weak $u$ solutions enjoying the integrability property $L^4(0,T;L^4(\OO))$. Actually, in \cite{GPGU} the author proves that any very weak solution $u\in L^4(0,T;L^4(\OO))$ satisfies an energy equality.       
\end{itemize}}\end{rem}
\begin{rem}\label{RT}{\rm\phantom\newline\begin{itemize}\item  It is well known the following result proved in \cite{CKN}:\vspace{3pt}\\\null\hskip0.5cm{\sl Assume $v_0\in J^{1,2}(\R^3)$. Then there exists a   $R(v_0)>0$ such that the corresponding suitable  weak solution\footnote{\,Here suitable weak solution is meant in the sense of Caffarelli-Kohn-Nirenberg}  $(v,\pi)$ is smooth in $(0,\infty)\times\R^3-B_R$, and  the pressure $\pi\in L^\frac43(0,\infty;L^2(\R^3))$\,.}\par Starting from the weak formulation, by means of the Prodi's arguments, it is not difficult to prove that for a such weak solution the following localized energy equality holds:
$$\dm h^\frac12v(t)\dm_2^2+2\!\intll0t\!\dm h^\frac12\n v(\tau)\dm_2^2d\tau=\dm h^\frac12 v_0\dm_2^2+\!\intll0t\!\big[(\Delta h,|v|^2)+(v\otimes v,\n h\otimes v)+2(\pi,\n h\cdot v)\big]d\tau\,, $$ for all $t>0$, and for all smooth non-negative function $h:\R^3\to\ov{\R}^+$ with support enclosed in $\R^3-B_R$. 
\item To date, with regard to the question of the validity of the energy equality \rf{EE}, we detect two different   ways of thinking. One looks for sufficient conditions in order to obtain the energy equality.  Another  tries to justify the inequality by means of turbulence arguments.\item
The former is based on  a wide literature originated\footnote{\,Prodi's result on the energy equality is in  Lemma\,2 of the quoted paper, which is devoted to the uniqueness of weak solutions enjoying an extra condition. The extra condition for the uniqueness is the same   used by Serrin in \cite{JS}, that is $L^{\frac{2p}{p-3}}(0,T;L^p(\OO))$,\,$p>3$.} by Prodi (1959) in \cite{GP}, that required the integrability property $L^4(0,T;L^4(\OO))$  to a weak solution $v$    (more properly an extra condition)  in order to obtain the energy equality for $v$.  
 \par In the setting of extra conditions (Prodi's kind),  we find some that  concern the derivatives of the weak solution. 
 The goal is to weaken  ``Prodi's condition''. \par Actually, in \cite{CC} the authors assume $v\in L^3(0,T;D(A^\frac5{12}_2))$ and in \cite{FT} the authors  assume $v\in L^3(0,T;D(A^{\frac 14}_{\frac{18}7}))$. \par In papers \cite {B,BC} and in Theorem\,1.3 Ch. I   of \cite{MJ} (PhD thesis), the   authors  assume $\n v\in L^r(0,T;L^s(\OO))$ with $\frac {2n}s+\frac{n+2}r=n+2$ ($n\geq3$ is the Euclidean dimension of $\OO$)\,\footnote{\, For $\beta\in(0,1)$, $A^{\beta}_q$ is the fractional power of the Stokes operator $A_q:=-P_q\Delta$. If $n=3$, due to variability of the exponents $r,s$, the extra assumptions  in the papers \cite{B,BC,MJ}  are not in all comparable with the ones of \cite{CC,FT}.}.  \par  However, the extra conditions, as a matter of course,  force the initial datum to be suitably more regular  than  the assumption  $v_0\in J^2(\OO)$. Therefore, these  results can only concern the weak solutions corresponding to a subset of   initial data in $J^2(\OO)$.  In this setting interesting results are obtained in \cite{KS,KOS}.\par Conversely, more  recent than the former, some papers have been devoted to study  the compatibility between the initial datum only in $L^2$   and  the validity of the energy equality of a weak solution, see \cite{CGM-I,CGM-II,CGM-III,Mi}. In particular, the result found in \cite{Mi} suggests that there is no  incompatibility\footnote{\,In \cite{Mi} it is proved that also the extra conditions for the regularity can be relaxed in such a way to be compatible  with the assumption $v_0$ just in $J^2(\OO)$. This fact points out that the characterization between the initial data and some extra-conditions of Serrin's kind, investigated by some authors, see {\it e.g.} \cite{KS,KOS}, concerns the only conditions and cannot be regarded as a characterization of the  regularity properties of a Leray-Hopf weak solution.\par In \cite{GFC}, from a different point of view, the extra condition $L^4(\vep,T;L^4(\OO))$ is deduced and applied  for a special weak solution. Consequently,  a local energy equality holds too.}  between  the extra   condition $L^4(\vep,T;L^4(\OO))$,\,$\vep>0$, which is enough to ensure the energy equality,  and the assumption  $v_0$ in $ J^2(\OO)$. \item Instead, the authors of \cite{C-I,C-II} support the following ideas. In the case of  ``less regularity'', the idea is that    the presence of turbulences in fluid motions causes an extra  dissipation, the one that balances the gap in the energy inequality \rf{NS-III}. Actually, they consider as possible an analogy between the gap in \rf{NS-III} and  the ``anomalous dissipation''  conjectured by Onsager\footnote{\,The Onsager conjecture is not part of the goals of this note. We refer to the fundamental results obtained in \cite{BDSV,CET,Sh-I}.} (1949) in \cite{Os} for weak solutions to the Euler equations.  In \cite{C-I,C-II}, the anomalous  dissipation $\vep[u^\nu](t,x)$, for all $\nu>0$ (viscosity coefficient), proposed in the energy relation is defined as   \be\label{CG-I}\vep[u^\nu](t,x):=\nu |\n u^\nu(t,x)|^2+ D[u^\nu(t,x)]\,,\ee with $D[u^\nu]\geq0$. So that, in place of \rf{NS-III} the energy relation becomes  \be\label{CG-II}\dm u^\nu(t)\dm_2^2+ 2\nu\intll 0t\dm \n u^\nu(\tau)\dm _2^2d\tau+2\intll0t\dm D[u^\nu](\tau)\dm_1d\tau=\dm u_0\dm_2^2\,,\mbox{\, a. e. in }t>0.\ee   In section\,3. of \cite{C-II}, the validity of \rf{CG-I} is supported by considering a Leray's approximation (see problem \rf{NSM} of this note).  In particular, denoted by $\{u^m\}$ the sequence of solutions to problem \rf{NSM},  by virtue of the properties of the functional convex and semicontinuous associated to the  dissipation, the authors achieve the following limit property (in the following $u$ is the weak solution, weak limit of $\{u^m\}$, the symbol $\nu$  is omitted): 
\be\label{DR}\lim_m|\n u^m(t,x)|^2-|\n u(t,x)|^2= D[u]\geq0\,.\ee Hence, if $D[u^\nu(t,x)]>0$ on a set of non-zero measure, the extra energy is achieved. But, employing the strong convergence stated in Lemma\,\ref{SCG}, that holds  for all the Leray sequences $\{u^m\}$, the limit property \rf{DR} can be only satisfied like equal to zero a.e. in $(t,x)$. So that, by means of \rf{DR} one cannot justify   formula \rf{CG-I}. Hence,   \rf{CG-II} is not justified.\end{itemize}}\end{rem}
 \par{\it Outline of the proof.}
\par Our result is an existence result, so that we suitably construct our weak solution $v$. For this aim we need to recall some  well known results, that we partially reproduce in order to make the note self-contained. \par The starting point is a sequence  $\{v^m\}$, whose $m$th element is a smooth  solution to the {\it mollified}  Navier-Stokes IBVP (see \rf{NSM}, this is the Leray approach). \par The first step is to prove that, for all $q\in[1,2)$, the sequence $\{\n v^m\}$ is strongly convergent in $L^q(0,T;L^2(\OO))$ (Lemma\,\ref{SCG}). 
Subsequently, by means of an auxiliary function, and making use of the energy relation for the smooth solutions $\{v^m\}$, we are able to prove a first result of convergence that leads to the following relation (formula \rf{LS-I}):
$$\sfrac2\pi\lim_{\alpha \to1^-}\sfrac 1{1-\alpha }\lim_m\intl{J(\alpha ,m)}\hskip-0.2cm\sfrac{\dm v^m\dm_2^2}{1+\dm \n v^m\dm_2^4} \sfrac d{d\tau}\dm \n v^m\dm_2^2d\tau=\dm v(s)\dm_2^2-\dm v(t)\dm_2^2-2\intll st\dm \n v\dm_2^2 d\tau\,,  $$
($J(\alpha,m)\subset (s,t)$, see \rf{LS-IO}, for all $(\alpha,m)$ the set $J(\alpha,m)$ is at most the union  of a countable family of pairwise disjoint     intervals). Thanks always to special properties of our auxiliary function and to the energy relation valid for the sequence $\{v^m\}$, we prove that the term of on the left-hand side is equal (formula \rf{SLM-VI}) to
$$ \sfrac2\pi\lim_{\alpha \to1^-}\sfrac1{1-\alpha } \lim_m\hskip-0.2cm\intl{J(\alpha ,m)}\hskip-0.25cm\sfrac{\dm v^m\dm_2^2}{1+\dm \n v^m\dm_2^4}\sfrac d{d\tau}\dm \n v^m\dm_2^2 d\tau=\lim_{\alpha \to1^-}\overline{\lim_m}\hskip-0.2cm\intl{J(\alpha ,m)}\hskip-0.3cm\dm \n v^m\dm_2^2d\tau=\lim_{\alpha \to1^-}{\underset m{\underline\lim}}\hskip-0.2cm\intl{J(\alpha ,m)}\hskip-0.3cm\dm \n v^m\dm_2^2d\tau\,.$$ 
Subsequently, we prove that the previous limits initially considered on $J(\alpha,m)$ ({\it a priori}, for all $(\alpha,m)$,   countable set of disjoint  open intervals) can be relaxed to a subset $J'(\alpha,m)$ of disjoint  open intervals whose cardinality is finite  (\texttt{card}$J'(\alpha,m)=$\texttt{n}$(\alpha)<\lfloor c\tan\alpha\frac\pi2\rfloor$, for all $m\in\N$). This property of finite  cardinality  allows us by means of further  arguments to prove  \rf{CTII-II}. 
\hfill $\Box$
\vskip0.2cm  
The plan of the paper is the following. After recalling and proving  some preliminary results in sect.\,\ref{SPL}, in sect.\,\ref{MNSS} we introduce the Navier-Stokes initial boundary value problem with a mollified  non-linear term, in order to work with a sequence $\{(v^m,\pi^m)\}$ of smooth approximating solutions, whose limit in ``metric of the energy'' gives our weak solution. 
Finally, in sect.\,\ref{PSEC} we furnish the proof of  Theorem\,\ref{CTI}. \vskip0.2cm\par{\bf Acknowledgement} - The author would like to thank     F. Crispo and   C.R. Grisanti with whom,  in the joint works \cite{CGM-I,CGM-II,CGM-III}, he shared and discussed with  interest the same question. \par Moreover, the author would like to thank a referee who, by bringing to the attention of the author several misprints, improved the reading of the paper.
\section{\label{SPL}Some preliminary lemmas} 
We start by recalling the $L^q$-Helmholtz decomposition, that is $$L^q(\OO)\equiv J^q(\OO)\oplus G^q(\OO)\,,$$ where $J^q(\OO)$ is the completion of $\mathscr C_0(\OO)$ in $L^q(\OO)$ and $G^q(\OO):=\{w\in L^q(\OO):w\equiv \n h \mbox{ with }h\in W_{\ell oc}^{1,q}(\OO)\}\,.$  By the symbol $P_q$ we denote the projection  from $L^q(\OO)$ onto $J^q(\OO)$. In the case of $q=2$, we just write $P$. For details on the Helmholtz decomposition see the monograph \cite{GPG}, and see also \cite{Sl,MRV} for a different approach to the problem that is due to Solonnikov.\par By the symbol $-P_q\Delta$ we mean the Stokes operator defined on $J^{1,q}(\OO)\cap W^{2,q}(\OO)$ with range $J^q(\OO)$. Here the symbol  $J^{1,q}(\OO)$ denotes the completion of $\mathscr C_0(\OO)$ in  $W^{1,q}(\OO)$. For further results on the Stokes operator we refer to the monograph \cite{GPG}. \begin{lemma}\label{GNM}{\sl There exists a constant $c>0$ such that \be\label{GNM-I}\dm w\dm_\infty \leq c\dm P\Delta w\dm_2^\frac12\dm \n w\dm_2^\frac12\,,\mbox{ for all }w\in J^{1,2}(\OO)\cap W^{2,2}(\OO)\,.\ee}\end{lemma}\bp Estimate \rf{GNM-I} is an inequality of the  Gagliardo-Nirenberg kind, whose right-hand side has the Stokes operator as max order of derivatives. For the proof see \cite{MRIII,MRIV}. \ep  \begin{lemma}[Friedrichs's lemma]\label{FR}{\sl Let $\OO$ be a bounded domain, and $\{a_p\}$ an orthogonal basis in $L^2(\OO)$. For all $\vep>0$ there exists  $N\in \N$ such that
$$\dm u\dm_2\leq (1+\vep)\big[\mbox{${\overset{N}{\underset{k=1}\sum}}$}(u,a^k)^2\big]^\frac12+\vep\dm\nabla u\dm_q,\mbox{ for all }u\in W^{1,q}(\OO)\,,$$ provided that $q>\frac65$. }\end{lemma}
\bp This lemma is a generalization of the well known Friedrichs's lemma stated for $q=2$. The proof is given in    \cite{LSU} Ch.II Lemma\,2.4.\ep
\begin{lemma}\label{WC}{\sl Let $\{h_m(t)\}\subset L^1(0,T)$ be a sequence of   non-negative functions such that $\dm h_m\dm_1\leq M<\infty$, for all $m\in\N$. Also, assume that $h_m(t)\to h(t)$ a.e. in $t\in (0,T)$ with $h(t)\in L^1(0,T)$.   Then we get,\be\label{WC-I}\displ\mbox{for all }\alpha\in(0,\alpha_0)\,,\quad\lim_m\intll0Th_m(t)p(\alpha,h_m(t))dt= \intll0Th(t)p(\alpha,h(t))dt\,,\ee and \be\label{WC-IO}\lim_{\alpha\to\widehat\alpha^-}\lim_m\intll0Th_m(t)p(\alpha,h_m(t))dt=\intll0Th(t)dt\,,\ee provided that, for all $\alpha\in(0,\widehat\alpha)$, the function $p(\alpha,r)$ is continuous in $r>0$ and 
$$p(\alpha,r):=\left\{\ba{ll}\!1\,,&\mbox{if }r\in[0,g(\alpha)]\,, \VS \!\mbox{is decreasing, with }\displ\lim_{r\to\infty}p(\alpha,r)=0\,, &\mbox{if }r>g(\alpha)\,,\ea\right.$$ where $g(\alpha)$ denotes a strictly increasing and continuous function with $\displ\lim_{\alpha\to\widehat\alpha^-}g(\alpha)=\infty$.
}\end{lemma}
\bp We have
$$\ba{ll}\displ\intll0Th_m(t)p(\alpha,h_m(t))dt\hskip-0.2cm&\displ=\intll0T(h_m(t)-h(t))p(\alpha,h_m(t))dt+
\intll0Th(t)p(\alpha,h_m(t))dt\\&=:I_1(\alpha,m)+I_2(\alpha,m)\,.\ea$$ Since $h_m(t)\to h(t)$ a.e. in $t\in(0,T)$, then, for all $\alpha\in(0,\widehat\alpha)$, $p(\alpha,h_m(t))\to p(\alpha,h(t))$ a.e. in $t\in (0,T)$. Since $p(\alpha,h_m(t))\leq1$ for all $t\in(0,T)$, recalling that $h(t)\in L^1(0,T)$, then the following limit holds: \be\label{CRGFIC-XII}\lim_mI_2(\alpha,m)=\intll0Th(t)p(\alpha,h(t))dt\,,\mbox{ for all }\alpha\in(0,\widehat\alpha)\,.\ee Now, we consider $I_1$ that, in our assumptions, is  bounded with a constant independent of $\alpha\in(0,\widehat\alpha)$ and $m\in\N$. Our goal is to prove that $\lim_mI_1(\alpha,m)=0$, for all $\alpha\in(0,\widehat\alpha)$. We point out that, for $\vep\in(0,\widehat\alpha-\alpha)$ and for all  $m\in\N$, $$(0,T)=\{t:h_m(t)\leq g(\widehat\alpha-\vep)\}\cup\{t:g(\widehat\alpha-\vep)<h_m(t)\}=:J_m^1(\vep)\cup J_m^2(\vep)\,.$$ Hence, for all $m\in\N$, we write
$$\ba{ll}\displ I_1(\alpha,m)\hskip-0.2cm&\displ=\intll0T\chi_{J^1_m}(t)(h_m(t)-h(t))p(\alpha,h_m(t)) dt+\intll0T\chi_{J_m^2}(t)(h_m(t)-h(t))p(\alpha,h_m(t))dt\VSE=I^1_1(\alpha,m,\vep)+I^2_1(\alpha,m,\vep)\,.\ea$$
Letting $m\to\infty$ we get  $\chi_{J^1_m}(t)|h_m(t)-h(t)|p(\alpha,h_m(t))\leq |h_m(t)-h(t)| \to 0$ a.e. in $t\in (0,T)$. Recalling that $p(\alpha,h_m(t))\leq1$, we have $|\chi_{J^1_m}(t)(h_m(t)-h(t))p(\alpha,h_m(t))|\leq g(\widehat\alpha-\vep)+h(t)$ for all $t\in (0,T)$. Hence, by virtue of Lebesgue's dominated convergence theorem, for all $\alpha\in(0,\widehat\alpha)$ and  for all $\vep\in(0,\widehat\alpha-\alpha)$, we arrive at 
\be\label{CRGFIC-X}\lim_mI_1^1(\alpha,m,\vep)=0\,.\ee
For the second integral we point out that $\chi_{J^2_m(t)}p(\alpha,h_m(t))\leq p(\alpha,g(\widehat\alpha-\vep))$ for all $t\in(0,T)$. Recalling that the sequence $\{h_m\}$ is bounded in $L^1(0,T)$ and that $h\in L^1(0,T)$,  we deduce 
\be\label{CRGFIC-XI}\Big|I_1^2(\alpha,m,\vep)\Big|\leq cp(\alpha,g(\widehat\alpha-\vep))\,,\ee with $c$ independent of $\alpha\in(0,\widehat\alpha)$, $\vep\in(0,\widehat\alpha-\alpha)$ and  $m\in\N$. Hence, via \rf{CRGFIC-X} and \rf{CRGFIC-XI}, for all $\alpha\in(0,\widehat\alpha)$ and $\vep\in(0,\widehat\alpha-\alpha)$, we arrive at $${\overline{\underset m\lim}}\,|I_1(\alpha,m)|\leq cp(\alpha,g(\widehat\alpha-\vep))\,.$$   By the assumptions,  $p(\alpha,r)$ tends to zero for $r\to\infty$, for all $\alpha\in(0,\widehat\alpha)$, and since $\vep\in(0,\widehat\alpha-\alpha)$ is arbitrary,   we arrive at $$\lim_mI_1(\alpha,m)=0\,.$$ The last one and the limit property \rf{CRGFIC-XII} prove \rf{WC-I}. \par Being $h(t)\in L^1(0,T)$ and $p(\alpha,\rho)\leq1$, and, letting $\alpha\to\widehat\alpha$, $ p(\alpha,\rho)\to1$,   via   Lebesgue's    dominated convergence theorem, then from \rf{WC-I} we get $$\lim_{\alpha\to\widehat\alpha} \intll0Th(t)p(\alpha,h(t))dt=\intll0Th(t)dt\,,$$ that proves \rf{WC-IO}. \ep 
\begin{rem}{\rm We stress the peculiarity of the lemma whose thesis is  connected to the auxiliary function $p$. Actually, for $p\equiv1$ the result fails to hold. We have the example of sequence $\{u_m\}$ with $u_m=m$ for $t\in(0,\frac1m)$ and $u_m=0$ for $t\in(\frac1m,1)$. Then $\dm u_m\dm_1=1$, for all $m\in\N$, $u_m\to0=u$ on $(0,1)$, but $\dm u_m\dm_1$ does not converge  to $ 0$.} \end{rem}
\begin{lemma}\label{CRGFIC-L}{\sl Let  $\{\rho_m\}$ be a sequence of measurable functions. Assume that, a.e. on $(s,t)$, $\{\rho_m\}$ converges  to $\rho\in L^1(s,t)$. Let $J(\alpha,m):=\{\tau\in(s,t) : \rho_m(\tau)>\tan\alpha\frac\pi2\}$. Assume that  for a constant $c$, independent of $\alpha\in(0,1)$, the following holds: \be\label{CRGFIC-O}|J(\alpha,m)|\leq  \sfrac c {\tan\alpha\frac\pi2 }\,,\mbox{ for all  }m\in \N\,.\ee   Then, we get
\be\label{CRGFIC-I}\lim_{\alpha\to1^-}\sfrac1{1-\alpha}\,\overline{\lim_m}\intl{J(\alpha,m)}\sfrac{\rho_m^2}{1+\rho_m^2\hskip-0.1cm}\hskip0.1cmd\tau =0\,.\ee}\end{lemma}
\bp We initially remark that the left-hand side of \rf{CRGFIC-I} is well posed for all $\alpha\in(0,1)$ and $m\in\N$. Actually,  $\rho_m^2/(1+\rho_m^2)\in(0,1)$ and, by virtue of assumption \rf{CRGFIC-O}, $ \displ{\underset{\alpha\to1^-}{\overline\lim}}\sfrac{|J(\alpha,m)|}{1-\alpha}\leq \lim_{\alpha\to1^-}\sfrac c{(1-\alpha)\tan\alpha\frac\pi2 }\leq c\sfrac\pi2\,.$ By means of a simple computation, we get
$$\ba{ll}\displ \overline{\lim_m}\hskip-0.2cm\intl{J(\alpha,m)}\hskip-0.2cm\sfrac{\rho_m^2}{1+\rho_m^2\hskip-0.1cm}\hskip0.1cmd\tau\hskip-0.2cm&\displ=\overline{\lim_m}\Big[\hskip-0.2cm\intl{J(\alpha,m)}\hskip-0.2cm\Big[\sfrac{\rho_m^2}{1+\rho_m^2\hskip-0.1cm}-\sfrac{\rho^2}{(1+\rho_m^2)^\frac12(1+\rho^2)^\frac12\hskip-0.1cm}\hskip0.1cm\Big]d\tau\\&\displ\hskip2cm+ \intl{J(\alpha,m)}\hskip-0.2cm\sfrac{\rho^2}{(1+\rho_m^2)^\frac12(1+\rho^2)^\frac12\hskip-0.1cm}\hskip0.1cmd\tau\Big]\\&\displ=:\overline{\lim_m}\,\Big[I_1(\alpha,m)+I_2(\alpha,m)\Big]\,.\ea$$For $I_1$ we have
$$\sfrac{\rho_m^2}{1+\rho_m^2\hskip-0.1cm}-\sfrac{\rho^2}{(1+\rho_m^2)^\frac12(1+\rho^2)^\frac12\hskip-0.1cm}\to0\,,\; \mbox{ a.e. in }\tau\in(s,t),$$
and$$\Big|\sfrac{\rho_m^2}{1+\rho_m^2\hskip-0.1cm}-\sfrac{\rho^2}{(1+\rho_m^2)^\frac12(1+\rho^2)^\frac12\hskip-0.1cm}\Big|\leq (1+\rho)\,,\;\mbox{ a.e. in }\tau\in (s,t)\,.$$
Hence, applying  Lebesgue's dominated convergence  theorem, we obtain $$\lim_mI_1(\alpha,m)= \intll st\Big[\sfrac{\rho_m^2}{1+\rho_m^2\hskip-0.1cm}-\sfrac{\rho^2}{(1+\rho_m^2)^\frac12(1+\rho^2)^\frac12\hskip-0.1cm}\Big]d\tau=0\,, \mbox{ for all }\alpha\in(0,1)\,.$$ 
Since $\rho_m> \tan\alpha\frac\pi2 $ for all $t\in J(\alpha,m)$,  
for the second integral we have$$I_2(\alpha,m)\leq \frac1{\tan\alpha\frac\pi2 \hskip-0.1cm}\hskip0.1cm\intl{J(\alpha,m)}\rho d\tau\,.$$ We set $$X(\alpha):=\{Y\subset (s,t):|Y|\leq \frac{c}{\tan\alpha\frac\pi2 \hskip-0.1cm}\hskip0.1cm\}\,.$$ Since in \rf{CRGFIC-O} we assumed $|J(\alpha,m)|\leq \frac{c}{\tan\alpha\frac\pi2\hskip-0.1cm}$\hskip0.1cm, then $J(\alpha,m)\in X(\alpha)$ and $$I_2(\alpha,m)\leq \sfrac1{\tan\alpha\frac\pi2 \hskip-0.1cm}\hskip0.1cm\sup_{Y\in X(\alpha)}\intl Y\rho d\tau\,.$$So that we get
$${\overline{\underset m\lim}}\hskip-0.2cm\intl{J(\alpha,m)}\hskip-0.2cm\sfrac{\rho_m^2}{1+\rho_m^2\hskip-0.1cm}\hskip0.1cmd\tau\leq 
\sfrac1{\tan\alpha\frac\pi2\hskip-0.1cm}\hskip0.1cm\sup_{Y\in X(\alpha)}\intl Y\rho d\tau\,.$$ Hence, we arrive at
$$\sfrac 1{1-\alpha}\hskip0.05cm{\overline{\underset m\lim}}\hskip-0.2cm\intl{J(\alpha,m)}\hskip-0.2cm\sfrac{\rho_m^2}{1+\rho_m^2\hskip-0.1cm}\hskip0.1cmd\tau\leq 
\sfrac 1{(1-\alpha)}\sfrac1{\tan\alpha\frac\pi2 \hskip-0.1cm}\hskip0.1cm\sup_{Y\in X(\alpha)}\intl Y\rho d\tau\,.$$
Being $\displ\lim_{\alpha\to1^-}(1-\alpha)\tan\alpha\frac\pi2 =\frac2\pi$, we are going to prove that \be\label{CRGFIC-II}\lim_{\alpha\to1^-}\sup_{Y\in X(\alpha)}\intl Y\rho d\tau=0\,.\ee In this way, the thesis of the lemma holds. By virtue of the theorem on the absolute continuity of the integral, for all $\vep>0$ there exists $\ov\alpha\in(0,1)$ such that\be\label{SSLLMM-O}\ba{l}\displ\mu_1(Y)<\sfrac c{\tan\ov\alpha\frac\pi2\hskip-0.1cm}\hskip0.01cm\;\Rightarrow\intl{Y}\rho d\tau<\vep\,,\\\displ\mbox{in particular, for all }\alpha\in(\ov\alpha,1):\mu_1(Y)\leq \sfrac c{\tan\alpha\frac\pi2 \hskip-0.1cm}\hskip0.01cm<\sfrac c{\tan\ov\alpha\frac\pi2 \hskip-0.1cm}\hskip0.01cm\;\Rightarrow\intl{Y}\rho d\tau<\vep\ea\ee For the same $\vep$, by virtue of the $sup$-property, we get the existence of $\ov Y\in X(\alpha)$ such that $$\sup_{Y\in X(\alpha)}\intl Y\rho d\tau<\intl {\ov Y}\rho d\tau+\vep<2\vep\,,$$ where in the last step we taken \rf{SSLLMM-O}$_2$ into account. Since $\vep$ is arbitrary,  we have proved  \rf{CRGFIC-II}.
\ep
 \section{\label{MNSS}The IBVP for the mollified  Navier-Stokes equations}
 We introduce an auxiliary  Navier-Stokes initial boundary value problem:
\be\label{NSM}\ba{cc}v_t^m+\mathbb J_m[v^m]\cdot\n v^m+\n\pi^m=\Delta v^m\,,&\n\cdot v^m=0\,,\mbox{ on }(0,T)\times \OO\,,\VS v^m=0\mbox{ on }(0,T)\times\po\,,& v^m=v^m_0\mbox{ on }\{0\}\times\OO\,,\ea\ee
where $\mathbb J_m[\,\cdot\,]\equiv J_{\frac1m}[\,\cdot\,]$ and $J_{\frac1m}[\,\cdot\,]$ is the Friedrichs (space) mollifier,  and $\{v_0^m\}\subset\mathscr C_0(\OO)$ converges to $v_0$ in $L^2$-norm. \begin{tho}\label{NSM-T}{For all $v_0^m\in \mathscr C_0(\OO)$ there exists a unique smooth solution, for $t>0$, to problem \rf{NSM} defined for all $T>0$. In particular, for all $T>0$, we get  \be\label{NSM-I}\ba{c}\mbox{for\,all\,}\varphi\!\in\! J^2(\OO), \,\{(v^m(t),\varphi)\}\!\subset\! C((0,T))\mbox{\,is\,uniformly\,equicontinuous\,and\,bounded\,sequence}\VS v^m\in C^1((0,T);J^{1,2}(\OO))\cap L^2(0,T;W^{2,2}(\OO))\,,\;v_t^m\in L^2(0,T;L^2(\OO))\,,\ea\ee and, for all $t>s\geq0$, 
\be\label{NSM-II}\dm v^m(t)\dm_2^2+2\intll st\dm\n v^m\dm_2^2d\tau= \dm v^m(s)\dm_2^2\leq \dm v_0\dm_2^2\,.\ee  Moreover,   the following estimate holds:
\be\label{NSM-III}\dm \n v^m(t)\dm_2^2+\sfrac12\intll\theta t\Big[ \dm P\Delta v^m\dm_2^2+\dm v^m_t\dm_2^2\Big]d\tau\leq (2c\dm  v_0\dm_2^2)^{-1}\,,\mbox{ for all }t\geq\theta\mbox{  and  } m\in\N,\ee where we have  $\theta=c\dm v_0\dm_2^4$ with  a constant $c$  independent of $t,m$ and $v_0$.}\end{tho}
\bp For the existence and the regularity of a solution $(v^m,\pi^m)$  we can employ the well known Faedo-Galerkin method as proposed in \cite{Hy,Tm} ({\it e.g.}, one founds developed this idea    in Appendix of  \cite{CKN} or in sect.\,2 of \cite{GM}). In particular one arrives at proving \rf{NSM-I}-\rf{NSM-II}.   \par We prove estimate\footnote{\, Estimate \rf{NSM-III} is generally given  on the weak solution  $v$, it furnishes the regularity of $v$ for $t\geq c\dm v_0\dm_2^4$. This is a result related to the Leray's partial regularity,  called structure theorem of a Leray's weak solution.  Here we furnish the proof on $\{v^m\}$. One gets the same regularity  for the  weak solution $v$ with the instant $\theta$ as endpoint of the regularity interval. } \rf{NSM-III}.  We apply the projection operator $P$ to equation \rf{NSM}$_1$, then, we consider the $L^2$-norm of both sides: $$\dm v_t^m-P\Delta v^m\dm_2^2=\dm P(J_m[v^m]\cdot\n v^m)\dm_2^2\,.$$ Since $(v^m_t,P\Delta v^m)=-\frac12\frac d{dt}\dm \n v^m\dm_2^2$, we get the following
\be\label{ALM-I}\sfrac d{dt}\dm\n v^m\dm_2^2+\dm P\Delta v^m\dm_2^2+\dm v^m_t\dm_2^2= \dm P(J_m[v^m]\cdot\n v^m)\dm_2^2\,.\ee
We estimate the right-hand side by means of \rf{GNM-I}:
\be\label{ALM-II}\dm P(J_m[v^m]\cdot\n v^m)\dm_2^2\leq \dm J_m[v^m]\cdot\n v^m\dm_2^2\leq \dm v^m\dm_\infty^2\dm \n v^m\dm_2^2\leq c\dm \n v^m\dm_2^3\dm P\Delta v^m\dm_2\,.\ee
Hence, we deduce
\be\label{NSM-IV}\sfrac d{dt}\dm\n v^m\dm_2^2+\frac12\dm P\Delta v^m\dm_2^2+\dm v^m_t\dm_2^2\leq c\dm \n v^m\dm_2^6\,.\ee From energy estimate \rf{NSM-II}, for all $m\in\N$, we deduce the existence of a $\theta_m\leq c\dm v_0\dm_2^4$ such that $\dm \n v^m(\theta_m)\dm_2^2\leq (2c\dm v_0\dm_2^2)^{-1}$. Actually, if the estimate  does not hold, we get
$$\sfrac12\dm v_0\dm_2^2=(2c\dm v_0\dm_2^{2})^{-1}c\dm v_0\dm_2^4<\intll0{c\dm v_0\dm_2^4}\dm\n v^m(\tau)\dm_2^2d\tau\leq\sfrac12\dm v^m_0\dm_2^2\leq\sfrac12\dm v_0\dm_2^2\,,$$ which is an {\it absurdum}. Since the differential energy relation furnishes $\dm \n v^m\dm_2^2\leq \dm v^m\dm_2\dm v^m_t\dm_2\leq \dm v_0\dm_2\dm v^m_t\dm_2$, from \rf{NSM-IV} we obtain$$\sfrac d{dt}\dm \n v^m\dm_2^2+\sfrac12\Big[\dm P\Delta v^m\dm_2^2+\dm v^m_t\dm_2^2\Big]<\dm\n v^m\dm_2^4 \dm v_0\dm_2^{-2}(2c\dm v_0\dm_2^2\dm \n v^m\dm_2^2-1)_{\big|t=\theta_m}\leq 0\,.$$This last, the bound for $\theta_m$ and $\dm \n v^m(\theta_m)\dm_2^2\leq (2c\dm v_0\dm_2^2)^{-1}$  easily lead to \rf{NSM-III}.\ep
\begin{rem}{\rm In place of $\mathbb J[\hskip0.03cm\cdot\hskip0.03cm]$ mollifier one can construct an approximating system     by means of the  Yosida Approximation. Then one arrives at the same result, that is,  estimates  \rf{NSM-I}-\rf{NSM-II} hold. For this result  we quote \cite{GS,MS}.}\end{rem}
\begin{lemma}\label{SCLD}{\sl For all $T>0$, the sequence $\{v^m\}$  furnished by Theorem\,\ref{NSM-T}   weakly converges to $v\in L^2(0,T;J^{1,2}(\OO))$. Moreover, there exists an extract, denoted again by $\{v^m\}$, such that,  for all $\varphi\in J^2(\OO)$, the sequence $\{(v^m(t),\varphi)\}$ converges, uniformly on $(0,T)$, to $(v(t),\varphi)\in C((0,T))$  and     converges strongly to $v$ in $L^2(0,T;L^2(\OO))$  }\end{lemma}
\bp The result is well known, it is a part of existence theorem of a weak solution enjoying the energy inequality in strong form. Thus,  we omit the details and  we limit ourselves to sketch the idea of the proof.   The weak convergence to $v$ is a consequence of the energy relation \rf{NSM-II}. Moreover, for all $\varphi\in J^2(\OO)$, by virtue of \rf{NSM-I}$_1$, any extract $\{(v^m(t),\varphi)\}$   converges uniformly on $(0,T)$, and, since by virtue of \rf{NSM-II}, for all $t>0$,  any extract $\{v^m\}$  admits $v$ as weak  limit in $J^{2}(\OO)$, then    $(v(t),\varphi)$ is a continuous function, {\it e.g. }see \cite{Ld}. This is enough to apply,   the Friedrichs Lemma\,\ref{FR} for the sequence $\{v^m\}\subset L^2(0,T;L^2(\OO))$ in the way suggested in \cite{Ld}, that is, for all $R>0$,
\be\label{FLL}\intll0T\dm v^m-v\dm_{L^2(\OO\cap B_R)}^2d\tau\leq (1+\vep)\mbox{$\underset{p=1}{\overset N\sum}$}\intll0T(v^m-v,a^p)^2d\tau+\vep \intll0T\dm \n v^m-\n v\dm_2^2d\tau\,. \ee
Hence,  the result of convergence  is immediate in the case of $\OO$ bounded. In the case of $\OO$ unbounded, in accord with  our assumptions, the result can be proved in the way proposed by Leray in \cite{L}. The  idea of the proof  is  to achieve for the sequence $\{v^m\}$ the conditions of the compactness theorem in $L^p(\R^n)$  for all $t>0$. Assume that the following holds:\be\label{LDD} \dm v^m(t)\dm_{L^2(|x|>R)}^2\leq  \dm v^m_0\dm_{L^2(|x|>\frac R2)}^2+ c(t)\psi(R) \,\mbox{ for any }t>0, \,R>2\ov R\mbox{ and }m\in\N\,,\ee 
with $c(t)\in L^\infty(0,T)$ and $\psi(R,v_0)=o(1)$. Set $$H(R,T):=T\dm v_0\dm_{L^2(|x|>R)}^2+\psi(R)\intll0Tc(t)dt +\intll0T\dm v(t)\dm_{L^2(|x|>R)}^2dt\equiv o(1)\,,$$   then, for all $T>0$, we get
$$\ba{ll}\displ\intll0T\dm v^m(t)-v(t)\dm_2^2dt\hskip-0.2cm&\displ= \intll0T\dm v^m(t)-v(t)\dm_{L^2(\OO\cap B_R)}^2dt+\intll0T\dm v^m(t)-v(t)\dm_{L^2(|x|>R)}^2dt\\&\displ
\leq 2\intll0T\dm v^m(t)-v(t)\dm_{L^2(\OO\cap B_R)}^2dt+2\dm v_0^m-v_0\dm^2_{L(|x|>\frac R2)}+2H(R,t)\,.\ea
$$Employing  \rf{FLL} for the first term on the right-hand side, and recalling that $\{v_0^m\}$ converges to $v_0$ in the  $L^2$-norm, letting  $m\to\infty$, the first two terms on the right-hand side tend  to zero. Secondary, being $H(R,T)=o(1)$,   letting $R\to\infty$, we get that the right-hand side  approaches zero. \par Estimate \rf{LDD} is proved in several papers concerning the question of the energy inequality \rf{NS-III}. For this reason  we do not give the proof, but, for the interested reader, we quote the proof furnished in  sect. 6.4.1 
of the paper \cite{CGM-I}.
\ep 
The following lemma was proved for the first time in \cite{CGM-I}
\begin{lemma}\label{SCG}{\sl Let   $\{v^m\}$ be the sequence furnished by Theorem\,\ref{NSM-T} and let $v$ be the limit ensured by Lemma\,\ref{SCLD}. Then, for all $q\in[1,2)$, the sequence $\{v^m\}$ strongly   converges to $v$   in $L^q(0,T;J^{1,2}(\OO))$.}\end{lemma}
\bp  In order to prove the strong convergence, we initially  prove that $\{P\Delta v^m\}$ is bounded in $L^\frac23(0,T;L^2(\OO))$\footnote{\, An analogous integrability property for  weak solutions has been obtained both in \cite{DF} and in \cite{FGT}. But our proof, very short, is original with regards to the ones of the quoted papers and is furnished on the sequence $\{v^m\}$, we are not interested to the property on the weak solution.}. We consider \rf{NSM-IV} again. 
Hence, we trivially deduce
$$\sfrac d{dt}\dm\n v^m\dm_2^2+\frac12\dm P\Delta v^m\dm_2^2+\dm v^m_t\dm_2^2\leq c\dm \n v^m\dm_2^6\leq c(1+\dm\n v^m\dm_2^2)^2\dm \n v^m\dm_2^2\,.$$
This last is integrated in the following way:
\be\label{DG-IV}\sfrac1{1+\dm\n v^m(0)\dm_2^2}+\intll0t\Big[\sfrac12\sfrac{\dm P\Delta v^m\dm_2^2+\dm v^m_t\dm_2^2}{(1+\dm \n v^m\dm_2^2)^2\hskip-0.15cm}\hskip0.15cm\Big]d\tau\leq \sfrac1{1+\dm\n v^m(t)\dm_2^2} +c\intll0t\dm\n v^m\dm_2^2d\tau\,.\ee Applying H\"older's reverse inequality with ``complementary'' exponents $q=-\frac12$ and $q'=\frac13$, uniformly in $m\in\N$, we get
$$\ba{ll}\displ\Big[\intll0t(\dm P\Delta v^m\dm_2^\frac23+\dm v_t^m\dm_2^\frac23)d\tau\Big]^3\hskip-0.2cm&\displ\leq \Big[\intll0t(1+\dm\n v^m\dm_2^2)d\tau\Big]^2\Big[1+c\intll0t\dm\n v^m\dm_2^2d\tau\Big]\VSE\leq \Big[t+\dm v_0\dm_2^2\Big]^2\Big[1+c\dm v_0\dm^2_2\Big]=: \frak A(t,\dm v_0\dm_2)\,,\mbox{ for all }t>0\,.\ea$$ For any pair $(m,p)$ and for all $t>0$, by means of an integration by parts and Holder's inequality, we obtain $$\ba{ll}\displ\intll0t\hskip-0.07cm\dm \n v^m\hskip-0.07cm-\n v^p\dm_2d\tau\!\leq \!\intll0t \hskip-0.07cm\dm P\Delta(v^m\hskip-0.07cm-v^p)\dm_2 ^\frac12
\dm v^m\hskip-0.07cm-v^p\dm_2^\frac12d\tau\hskip-0.25cm&\displ\leq\! \Big[\intll0t\hskip-0.07cm \dm P\Delta(v^m\hskip-0.07cm-v^p)\dm_2^\frac23\Big]^\frac34 \hskip-0.05cm\Big[\intll0t\hskip-0.07cm\dm v^m\hskip-0.07cm-v^p\dm_2^2d\tau\Big]^\frac14\VSE\leq2^\frac12 \frak A(t,\dm v_0\dm_2)^\frac1{4}\Big[\intll0t\dm v^m\hskip-0.07cm-v^p\dm_2^2d\tau\Big]^\frac14\,.\ea$$ By virtue of Lemma\,\ref{SCLD},
we obtain the Cauchy condition for $\{ v^m\}$ in $L^1(0,T;J^{1,2}(\OO))$. Via \rf{NSM-II}, the sequence $\{ v^m\}$ is bounded in $L^2(0,T;J^{1,2}(\OO))$. Hence, by interpolation we realize  the Cauchy condition in $L^q(0,T;J^{1,2}(\OO))$, for all $q\in[1,2)$. So that, the sequence admits the strong limit in $L^q(0,T;J^{1,2}(\OO))$, for all $q\in[1,2)$, which coincides with $v\in L^2(0,T;J^{1,2}(\OO))$.  The lemma is proved.
\ep
\begin{coro}\label{CSB}{\it  The sequence $\{v^m\}$ of solutions   to problem \rf{NSM}  converges      to $v$   in $J^{1,2}(\OO)$ uniformly on $t\geq\theta$\,. }\end{coro} \bp By virtue of Lemma\,\ref{SCLD} and Lemma\,\ref{SCG} we have that $\{v^m\}$   strongly converges to $v$   in $J^{1,2}$ a.e. in $t>0$. Without invalidating the thesis, we can assume that the strong convergence holds for $t=\theta$.  We can state the following estimates with a bound $\frak D$ independent of $m,\,t$ and $v$, and whose value is inessential for our aims;\begin{itemize}\item[]employing    estimates \rf{GNM-I}, \rf{NSM-II} and \rf{NSM-III}   we get
\be\label{CSB-II}\ba{ll}\displ\intll\theta t\dm v^m(\tau)\dm_\infty^2d\tau\hskip-0.2cm&\displ\leq c\Big[\intll\theta t\dm P\Delta v^m(\tau)\dm_2^2d\tau\Big]^\frac12\Big[\intll\theta t\dm \n v^m(\tau)\dm_2^2d\tau\Big]^\frac12=:\frak D_1\,,\\&\hskip0.2cm \mbox{ for all }m\in\N\mbox{ and }t>\theta\,;\ea\ee\item[]employing estimates \rf{NSM-II} and \rf{NSM-III} , we get\be\label{DQ}\ba{ll}\displ\intll\theta t\dm \n v^m(\tau)\dm_2^4d\tau\hskip-0.2cm&\displ\leq \max_{[\theta,t]}\dm \n v^m(\tau)\dm_2^2\intll\theta t\dm \n v^m(\tau)\dm_2^2d\tau=:\frak D_2\,,\\&\mbox{ for all }m\in\N\mbox{ and }t>\theta\,.\ea\ee\end{itemize} We denote by  $\frak D:=\max\{\frak D_1,\frak D_2\}$. Set $w:=v^m-v^p$, from system  \rf{NSM} by difference, we deduce
\be\label{CSB-III}\sfrac d{dt}\dm \n w\dm_2^2+\dm P\Delta w\dm_2^2+\dm w_t\dm_2^2=\dm P(v^m\cdot\n w)+P(w\cdot\n v^p)\dm_2^2\,\mbox{ for all }t>\theta\,.\ee Applying H\"older's inequality to the term of the right-hand side, we get
$$\ba{ll}\dm P(v^m\cdot\n w)+P(w\cdot\n v^p)\dm_2^2\hskip-0.2cm&\leq 2(\dm v^m\dm_\infty^2\dm\n w\dm_2^2+\dm w\dm_\infty^2\dm\n v^p\dm_2^2)\VSE\leq c(\dm v^m\dm_\infty^2\dm\n w\dm_2^2+\dm \n w\dm_2\dm P\Delta w\dm_2\dm \n v^p\dm_2^2)\,,\ea$$  where in the last estimate we employed \rf{GNM-I} again. Increasing the right-hand side of \rf{CSB-III} via the last estimate, and employing the Young inequality, we deduce the differential equation
$$\sfrac d{dt}\dm \n w\dm_2^2\leq c\dm \n w\dm_2^2(\dm v^m\dm_\infty^2+\dm \n v^p\dm_2^4)\,,\mbox{ for all }t>\theta\,.$$ Hence, by means of estimates  \rf{CSB-II} and \rf{DQ},  via an  integration, we arrive at 
\be\label{SCG-I}\ba{ll}\dm \n v^m(t)-\n v^p(t)\dm_2\hskip-0.2cm&=\dm \n w(t)\dm_2\leq \hskip0.05cm\exp[\frak D]\hskip0.05cm\dm \n w(\theta)\dm_2\VSE\leq \hskip0.05cm\exp[\frak D]\hskip0.05cm\dm \n v^m(\theta)-\n v^p(\theta)\dm_2\,,\mbox{ for all }m,p\in\N\mbox{ and }t\geq\theta\,.\ea\ee In the case of the $L^2$-norm, for all $t>\theta$, we consider the energy relation related to $w$:
$$\ba{ll}\frac d{dt}\dm w(t)\dm_2^2+2\dm \n w(t)\dm_2^2\hskip-0.2cm&=(w\cdot\n w,v^m)\leq \dm w\dm_6\dm \n w\dm_2\dm v^m\dm_3\leq \dm w\dm_6\dm \n w\dm_2\dm v^m\dm_2^\frac12\dm\n v^m\dm_2^\frac12\VSE\leq c\dm v_0\dm_2^\frac12\dm \n w\dm_2^2\dm \n v^m\dm_2^\frac12\leq c\dm v_0\dm_2^\frac12\dm \n w\dm_2^\frac12\Big[\dm \n v^m\dm_2^2+\dm \n v^p\dm_2^\frac32\dm\n v^m\dm_2^\frac12\Big] ,\ea$$ where we have increased by means of H\"older's inequality, Sobolev inequality and the Gagliardo-Nirenberg inequality.   Taking estimate \rf{NSM-II} into account, an integration on $t> \theta$,   furnishes
\be\label{SCLD-I}\ba{ll}\dm v^m(t)-v^p(t)\dm_2^2\hskip-0.25cm&=\dm w(t)\dm_2^2\leq \dm w(\theta)\dm_2^2+c\exp[\frak D]\dm v_0\dm_2^\frac52\hskip0.05cm\dm \n v^m(\theta)-\n v^p(\theta)\dm_2^\frac12\VSE \leq \dm v^m(\theta)-v^p(\theta)\dm_2^2+c\exp[\frak D]\dm v_0\dm_2^\frac52 \hskip0.05cm\dm \n v^m(\theta)-\n v^p(\theta)\dm_2^\frac12,\mbox{\,for all\,}m,p\in\!\N.\ea\ee  Now, from our assumption of strong convergence of $\{v^m(\theta)\}$ in $J^{1,2}(\OO)$, via estimates \rf{SCG-I} and \rf{SCLD-I},  the thesis of the corollary holds.
\ep
\begin{lemma}\label{GPM}{\sl The sequence $\{v^m\}$  furnished by Theorem\,\ref{NSM-T} enjoys the following estimate:\be\label{DG-III}\intll0t\sfrac1{(1+\dm \n v^m\dm_2^2)^2}\Big|\sfrac d{dt}\dm \n v^m\dm_2^2\Big|d\tau\leq 1+c\dm v_0\dm_2^2\,,
 t>0\,.\ee }\end{lemma}
\bp
From \rf{ALM-I} we are able to get
\be\label{ALM-III}\ba{ll}\displ\sfrac1{(1+\dm \n v^m\dm_2^2)^2}\Big|\sfrac d{dt}\dm \n v^m\dm_2^2\Big|\hskip-0.2cm&\displ=\Big|\sfrac{\dm P( J_m[v^m]\cdot\n v^m)\dm_2^2}{(1+\dm \n v^m\dm_2^2)^2}-\sfrac{\dm P\Delta v^m\dm_2^2+\dm v^m_t\dm_2^2}{(1+\dm \n v^m\dm_2^2)^2}\Big|\VSE
\leq c\sfrac{\dm\n v^m\dm_2^3\dm P\Delta v^m\dm_2}{(1+\dm \n v^m\dm_2^2)^2}+\sfrac {\dm P\Delta v^m\dm_2^2+\dm v^m_t\dm_2^2}{(1+\dm \n v^m\dm_2^2)^2}\VSE\leq \dm \n v^m\dm_2^2+c\sfrac {\dm P\Delta v^m\dm_2^2+\dm v^m_t\dm_2^2}{(1+\dm \n v^m\dm_2^2)^2}\,,\ea\ee
where estimating the right hand-side we have taken \rf{ALM-II} into account and applied the Young  inequality. Integrating \rf{ALM-III} on $(0,t)$, via estimate \rf{DG-IV}   and  via  energy relation \rf{NSM-II}, we arrive at \rf{DG-III}. 
\ep
\section{\label{PSEC}Proof  of Theorem\,\ref{CTI} and Corollary\,\ref{CoroC}}
\subsection{Existence of a weak solution $v$ to problem \rf{NS}, Theorem\,\ref{CTII}}
We start by proving the  existence of a weak solution. 
\begin{lemma}\label{EX-T}{\sl For all $v_0\in J^2(\OO)$ there exists a sequence $\{v^m\}$,   enjoying the following limit properties: \be\label{EST-O}\ba{ll}\mbox{for all }T>0,\hskip-0.1cm &\{v^m\}\rightharpoonup \hskip0.03cmv\mbox{ in }L^2(0,T;J^{1,2}(\OO))\,,\VSE   \{v^m\}\to v \mbox{ in }L^2(0,T;L^2(\OO))\cap L^q(0,T;J^{1,2}(\OO)), \,q\in[1,2)\,,\VS&\{v^m\}\to v\mbox{\,in\,}J^{1,2}(\OO),\mbox{\,a.e.\,in\,}t\!\in\![0,\theta)\mbox{\,and\,for\,all\,}t\geq\theta,\mbox{\,with\,}\theta\leq c\dm v_0\dm_2^4\,, \ea\ee where the constant $c$ is independent of    $v_0$, and the limit $v$ is a weak solution  to problem \rf{NS} that   enjoys  the  properties:
\be\label{EXT-I}\hskip-0.1cm\ba{c}\displ\dm v(t)\dm_2^2+2\!\intll st\!\dm \n v\dm_2^2d\tau\leq \dm v(s)\dm_2^2,\mbox{\,for\,all\,}t>s,\mbox{\,and\,for\,all\,}s\!\in\! [0,\infty)-I,\,I\hskip-0.04cm\subset\!(0,\theta),\,\mu_1(I)=0\,,\\\displ \mbox{for\,all\,}s\!\in\! I^C,\lim_{t\to s^+}\dm v(s)-v(t)\dm_2=0\,,\VS\mbox{for\,all\,}\varphi\!\in\! J^2(\OO),\,  (v(t),\varphi)\mbox{\,is\,a\,continuous\,function\,of }t\,.\ea\ee }\end{lemma}
 \bp The existence result of $v$   
is achieved following the approach furnished  by Leray in \cite{L}.
Actually, we consider the sequence stated in Theorem\,\ref{NSM-T} and its  limit $v$ stated in Lemma\,\ref{SCLD}, Lemma\,\ref{SCG}. Then the existence of $\theta$ and the related estimate are deduced from Theorem\,\ref{NSM-T}.  For all $t\in [0,T)$, $v$ is  also a weak  limit in $J^{2}(\OO)$ with $(v(t),\varphi)\in C([0,T))$, for all $\varphi\in J^2(\OO)$, that is \rf{EXT-I}$_3$.\par One easily verifies that the limit $v$ is a weak solution.   
\par Property \rf{EST-O}$_3$ is deduced from the strong convergence stated in Lemma\,\ref{SCG} and the regularity of $v$ achieved for $t>\theta$ in Corollary\,\ref{CSB}.
\par Estimate \rf{EXT-I}$_1$ is the so called energy inequality in strong form introduced by Leray. In order to deduce \rf{EXT-I}$_1$, we consider an instant $s\geq0$ in which   $\lim_m\dm v^m(s)\dm_2=\dm v(s)\dm_2$. In the case of $s\ne0$, the existence is ensured, almost everywhere in $s\in(0,\theta)$, by Lemma\,\ref{SCLD}, and for all $s\geq\theta$ by Corollary\,\ref{CSB}. We denote by $I$ the set of the possible instants $s$ for which the strong convergence fails to hold. By our arguments, $I\subset (0,\theta)$ and $\mu_1(I)=0$. Then, for all $s\in I^C$, we consider \rf{NSM-II}, and we perform the lower  limit of the left-hand side and the limit of the right-hand side of \rf{NSM-II}. Then one easily achieves the energy inequality for the weak limit $v$ as stated in \rf{EXT-I}. \par Properties \rf{EXT-I}$_2$ are an immediate consequence of the weak limit properties and of \rf{EXT-I}$_1$. \ep 
\subsection{A first ``partial energy equality'' for the weak solution}
 The first our goal is to prove that the weak solution $v$ furnished in  Lemma\,\ref{EX-T} enjoys the energy equality \be\label{EE-N}\ba{ll}\displ\dm v(t)\dm_2^2+2\!\intll0t\dm\n v(\tau)\dm_2^2d\tau-\dm v(s)\dm_2^2\hskip-0.25cm&=\displ2\!\lim_{\alpha \to1^-}\!{\lim_{\;\,m^{\null_{''}}}}\!\!\intl{\widetilde J(\alpha ,m^{\null_{''}})}\!\!\!\rho_{m''}d\tau =2\lim_{\alpha \to1^-}\!{\lim_{\;m^{\null_{''}}}}\!\mbox{$\underset{k=1}{\overset{\mbox{\scriptsize\rm\texttt{n}}(\alpha,m^{\null_{''}})}\sum}$}\!\!\intll{s_k}{t_k}\!\rho_{m''}d\tau\VS\dm v(t)\dm_2^2+2\!\intll0t\dm\n v(\tau)\dm_2^2d\tau-\dm v(s)\dm_2^2\hskip-0.25cm&\displ=2\lim_{\alpha \to1^-}{\lim_{\;m^{\null_{'}}}}\intl{\widetilde J(\alpha ,m^{\null_{'}})}\rho_{m'}d\tau\displ =2\lim_{\alpha \to1^-}\!{\lim_{\;m^{\null_{'}}}}\!\mbox{$\underset{k=1}{\overset{\mbox{\scriptsize\rm\texttt{n}}(\alpha,m^{\null_{'}})}\sum}$}\!\!\intll{s_k}{t_k}\!\rho_{m'}d\tau,\ea\ee where $\{\rho_{m''}\}$ and $\{\rho_{m'}\}$ are two suitable subsequences of $\{\rho_m\}$ and $\texttt{n}(\alpha,m'')$ and $\texttt{n}(\alpha,m')$ sequences of integers with upper bound $\lfloor 2c\tan\alpha\frac\pi2\rfloor$. \par For this aim, we are going to prove that some strong convergences hold. These convergences are a consequence of the strong convergences furnished in sect.\,2 and in sect.\,3, of some auxiliary functions and of the differential energy equality \rf{NSM-II} deduced for the elements of the sequence $\{v^m\}$. 
\par We denote by  $$\mathcal T:=\{s\geq 0: \{v^m(s,x)\} \mbox{ is strongly convergent to }v\mbox{ in }J^{1,2}(\OO)\}.$$
By virtue of \rf{EST-O}$_3$, the set $\mathcal T$ is not empty, and $$\mu_1([0,\infty)-\mathcal T)=\mu_1([0,\theta)-\{\mathcal T\cap [0,\theta)\})=0 .$$  \begin{lemma}\label{I-LS}{\sl Let $\{v^m\}$ be the sequence of solutions to \rf{NSM} and $v$ the  weak solution to \rf{NS} furnished in Lemma\,\ref{EX-T}. Let $t,s\in\mathcal T$.  
Then
 there exists   a family  $\{J(\alpha,m)\}$ of sets, where, for all $\alpha\in(\alpha_1,1)$, $\alpha_1>0$, and $m\geq m_0$, \be\label{LS-IO}\ba{l}J(\alpha ,m)\!:=\!{\underset{h\in\N(\alpha ,m)}\cup}\!(s_h,t_h),\,\N(\alpha,m)\mbox{\,at\,most\,countable set,\,and\,}|J(\alpha ,m)|\leq \!{\dm v_0\dm_2^2}\big[{\tan\alpha \sfrac\pi2}\big]^{-1},\VS\mbox{for all }h\in \N(\alpha ,m),\;(s_h,t_h)\!:=\big\{\tau:\tan\alpha \sfrac\pi2<\dm \n v^m(\tau)\dm_2^2\leq A_m\big\},\,A_m:=\max_{[s,t]}\dm \n v^m\dm_2^2\,,\VS\mbox{with }\dm \n  v^m(s_h)\dm_2^2=\dm \n v^m(t_h)\dm_2^2=\tan \alpha \sfrac\pi2 \,, \mbox{ and,  for }h\ne k,\,(s_h,t_h)\cap (s_k,t_k)=\emptyset\,.  \ea\ee  Set \be\label{HFH}h(\alpha,m):=\intl{J(\alpha ,m)}\hskip-0.2cm\sfrac{\dm v^m\dm_2^2}{1+\dm \n v^m\dm_2^4\hskip-0.1cm}\hskip0.1cm \frac d{d\tau}\dm \n v^m\dm_2^2d\tau\,,\ee for all $\alpha\in(\alpha_1,1)$ there exists  \be\label{LS-OI}\lim_mh(\alpha,m)\in[0,\sfrac12\dm v_0\dm_2^2)\,,\ee and the weak solution $v$ enjoys the following 
 special ``energy  relation'':
\be\label{LS-I}\sfrac2\pi\lim_{\alpha \to1^-}\sfrac 1{1-\alpha }\lim_m h(\alpha,m)=\dm v(s)\dm_2^2-\dm v(t)\dm_2^2-2\intll st\dm \n v\dm_2^2 d\tau\,.  \ee}\end{lemma}
\bp  We prove the claims \rf{LS-IO}-\rf{LS-I} by means of a suitable construction.  We set $$\alpha\in(0,1)\,,\;\rho>0\,,\;\;A(\rho):=\frac{\frac\pi2-\arctan\rho}{(1-\alpha)\frac\pi2}\,.$$ We define the function $$p(\alpha ,\rho):=\left\{\ba{ll}1,&\mbox{if }\rho\in [0,\tan\alpha \frac\pi2]\,,\vspace{4pt}\\ A(\rho),&\mbox{if }\rho\in(\tan\alpha \frac\pi2\,,\infty)\,.\ea\right.$$
Setting  $g(\alpha )=\tan\alpha \frac\pi2$, the function $p(\alpha ,\rho)$  enjoys the same properties of the
function  introduced in Lemma\,\ref{WC} with $\widehat\alpha:=1$. We consider the sequence $\{v^m\}$ of solutions. For all $m\in\N$, the energy equation \rf{NSM-II}, that for the convenience of the reader we reproduce, holds:
\be\label{ER}\sfrac d{dt}\dm v^m(t)\dm_2^2+2\dm \n v^m(t)\dm_2^2=0\;\Longleftrightarrow \;\dm v^m(t)\dm_2^2+2\intll st\dm \n v^m\dm_2^2d\tau=\dm v^m(s)\dm_2^2\,.\ee
We set $\rho_m(t):=\dm\n v^m(t)\dm_2^2$\,,
and we consider $p(\alpha ,\rho_m(t))$.
\par Since $s,t\in \mathcal T$\,, there exists  $\alpha _1$   such that $$\max\{\dm \n v(s)\dm_2^2,\dm\n v(t)\dm_2^2\}<\tan\alpha \sfrac\pi2\,,\mbox{ for all }\alpha \in(\alpha _1,1)\,,$$  where $v$ is the weak solution stated in Lemma\,\ref{EX-T}. Hence, by virtue of the   convergence in $J^{1,2}(\OO)$-norm stated in \rf{EST-O}$_3$, we state the existence of $m_0$ such that \be\label{GUB}\max\{\dm \n v^m(s)\dm_2^2,\dm\n v^m(t)\dm_2^2\}<\tan\alpha\sfrac \pi2\,, \mbox{ for all }m\geq m_0\mbox{ and }\alpha \in(\alpha _1,1)\,.\ee  
We denote by $J(\alpha ,m):=\{\tau\in(s,t):\rho_m(\tau)\in(\tan\alpha \frac \pi2,A^m]\}$. \par If $A_m\leq \tan\alpha \frac\pi2$, then $J(\alpha ,m)$ is an empty set. \par If $A_m>\tan\alpha\frac\pi2 $ holds, since  $\max\{\rho_m(s),\rho_m(t)\}<\tan \alpha \frac\pi2$, then one gets $(s,t)-J(\alpha,m)\ne\emptyset$\,. For $\tau\in(s,t)$,   the intersection between the straight line $(\tau,\tan\alpha\frac\pi2)$ and the curve $(\tau,\rho_m(\tau))$, by the continuity of $\rho_m$, detects the endpoints of open non empty intervals. Actually, 
since $\rho_m(s)<\tan\alpha\frac\pi2$ and $\rho_m(\tau)$ is continuous, there exists $\ov\tau$ such that $\rho_m(\ov\tau)=\tan\alpha\frac\pi2$ and $\rho_m(\tau)>\tan\alpha\frac\pi2$ for some right-neighborhood $U^+(\ov\tau)$. Analogously, being $\rho_m(t)<\tan\alpha\frac\pi2$, by the continuity there exists $\ov{\ov\tau}$ such that $\rho_m(\ov{\ov\tau})=\tan\alpha\frac\pi2$ and  $\rho_m(\ov{\ov\tau})>\tan\alpha\frac\pi2$ in   some left-neighborhood $U^-(\ov{\ov\tau}) $\,. 
A pair of distinct and successive    points $\ov\tau$ and $\ov{\ov\tau}$ are the endpoints of an open not empty interval $U\subseteq J(\alpha,m)$\,.
 Being 
 $U\ne\emptyset$ and open, $J(\alpha,m)$ is at most the union  of a countable family of pairwise disjoint    intervals $(s_h,t_h)$ with $\rho(s_h)=\tan \alpha\frac\pi2$ and $\rho(t_h)=\tan\alpha\frac\pi2\,$.  We denote by $\N(\alpha,m)$ the set of indexes $h$ whose cardinality is less  than or equal $\aleph_0$.\par    For the measure of $J(\alpha ,m)\subset(s,t)$, we get\be\label{MJ}\mbox{$\underset {h\in \N(\alpha,m)}\sum$}\!\!(t_h-s_h)\tan\alpha\sfrac\pi2=|J(\alpha ,m)|\tan\alpha \sfrac\pi2 \leq \!\!\intl{J(\alpha ,m)}\!\!\!\rho_m(\tau)d\tau\leq \intll st\rho_m(\tau)d\tau\leq\sfrac12\dm v_0\dm_2^2\,,\ee where we taken the energy relation \rf{ER} into account. Estimate \rf{MJ} completes \rf{LS-IO}. Recalling the definition of $p(\alpha,\rho_m(t))$, we have $$\sfrac d{d\tau}p(\alpha ,\rho_m(\tau))=\left\{\hskip0.15cm\ba{ll}0&\mbox{a.e. in }\tau\in (s,t)-\ov{J(\alpha,m)}\,,\VS\hskip-0.35cm-\sfrac2\pi\sfrac1{1-\alpha }\sfrac{\dot\rho_m(\tau)}{1+\rho_m^2(\tau)\hskip-0.1cm}\hskip0.1cm&\mbox{for all  }\tau\in J(\alpha ,m)\,,\ea\right.$$ where taken into account that, for all $\alpha \in(0,1)$, the function $p$ is a Lipschitz's function in $\rho_m$, and $\rho_m(t)\in C^1([s,t])$. Hence,    $p(\alpha ,\rho_m(t))$ is a Lipschitz's function in $t$.
We multiply equation \rf{ER} by  $p(\alpha ,\rho_m(t))$ and we integrate by parts on $(s,t)$:
\be\label{SLM-I} \sfrac2\pi\sfrac1{1-\alpha }\intl{J(\alpha ,m)}\hskip-0.2cm\sfrac{e_m}{1+\rho_m^2\hskip-0.1cm}\hskip0.03cm{\dot\rho_m(\tau)}d\tau=e_m(s)-e_m(t)-2\intll st\rho_m\hskip0.05cmp(\alpha ,\rho_m)d\tau\,,\ee where we set $e_m:=\dm v^m\dm_2^2$\,. Since the hypotheses of Lemma\,\ref{WC} are satisfied, and since  $s,t\in\mathcal T$,  by virtue of \rf{WC-I}, letting $m\to\infty$, each term on the right-hand side of \rf{SLM-I} admits limit. So that we arrive at 
\be\label{SLM-II} \sfrac2\pi\sfrac1{1-\alpha }\lim_m\intl{J(\alpha ,m)}\hskip-0.2cm\sfrac{e_m}{1+\rho_m^2(\tau)}\hskip0.03cm{\dot\rho_m(\tau)}d\tau=e(s)-e(t)-2\intll st\rho\hskip0.05cm p(\alpha ,\rho)d\tau\,,\ee where we set $e:=\dm v\dm_2$ and $\rho:=\dm \n v\dm_2^2$  ($v$ weak solution). The relation \rf{SLM-II} proves the limit property \rf{LS-OI}.
Hence, letting $\alpha \to1^-$, via \rf{WC-IO},  we arrive at
$$\sfrac2\pi\lim_{\alpha \to1^-}\sfrac 1{1-\alpha }\lim_m\intl{J(\alpha ,m)}\hskip-0.2cm\sfrac{e_m}{1+\rho_m^2}\hskip0.03cm{\dot\rho_m}d\tau=e(s)-e(t)-2\intll st\rho d\tau\,,$$
that, via the position \rf{HFH},  is equivalent to \rf{LS-I}.\ep
\begin{lemma}\label{CDD}{\sl Under the  assumptions of Lemma\,\ref{I-LS}, we get \be\label{SLM-VI} \sfrac2\pi\lim_{\alpha \to1^-}\sfrac1{1-\alpha }\lim_m\intl{J(\alpha ,m)}\sfrac{e_m}{1+\rho_m^2}\dot \rho_md\tau=2\lim_{\alpha \to1^-}\overline{\lim_m}\intl{J(\alpha ,m)}\rho_md\tau=2\lim_{\alpha \to1^-}{\underset m{\underline\lim}}\intl{J(\alpha ,m)}\rho_md\tau\,.\ee}\end{lemma}\bp   Recalling \rf{LS-IO} related to $J(\alpha ,m)$ given in Lemma\,\ref{I-LS}, by means of an integration by parts, we get $$\ba{ll}\displ \intl{J(\alpha ,m)}\!\!\!\sfrac{e_m}{1+\rho_m^2}\dot\rho_md\tau\hskip-0.2cm\!&\displ=\!\mbox{\large$\underset{h\in\N(\alpha ,m)}\sum$}\intll{s_h}{t_h}\sfrac{e_m}{1+\rho_m^2}\dot\rho_md\tau\\&\displ=\sfrac{\tan\alpha \frac\pi2}{1+\tan^2\!\alpha \frac\pi2\hskip-0.1cm}\hskip0.1cm\mbox{\large$\underset{h\in\N(\alpha ,m)}\sum$}\!\big[ e_m(t_h)-e_m(s_h)\big]-\!\mbox{\large$\underset{h\in\N(\alpha ,m)}\sum$}\intll{s_h}{t_h}\sfrac{\dot e_m\rho_m}{1+\rho_m^2}d\tau\\&\displ\hskip2cm +2\!\mbox{\large$\underset{h\in\N(\alpha ,m)}\sum$}\intll{s_h}{t_h}\sfrac{e_m\rho_m^2}{(1+\rho_m^2)^2\hskip-0.1cm}\hskip0.1cm\dot\rho_md\tau\,.\ea$$ Hence, via the energy relation \rf{ER}, we arrive at
$$\ba{l}\displ-\mbox{\large$\underset{h\in\N(\alpha ,m)}\sum$}\intll{s_h}{t_h}\sfrac{e_m}{1+\rho_m^2}\dot\rho_md\tau+2\mbox{\large$\underset{h\in\N(\alpha ,m)}\sum$}\intll{s_h}{t_h}\sfrac{e_m\rho_m^2}{(1+\rho_m^2)^2\hskip-0.1cm}\hskip0.1cm\dot\rho_md\tau\\\hskip6cm\displ=\sfrac{2\tan\alpha \frac\pi2}{1+\tan^2\!\alpha \frac\pi2\hskip-0.1cm}\hskip0.1cm\mbox{\large$\underset{h\in\N(\alpha ,m)}\sum$}\intll{s_h}{t_h}\rho_md\tau\;-2\!\!\mbox{\large$\underset{h\in\N(\alpha ,m)}\sum$}\intll{s_h}{t_h}\sfrac{\rho_m^2}{1+\rho_m^2}d\tau\,. \ea$$ Since $-\frac1{1+\xi^2\hskip-0.1cm}\hskip0.1cm+\frac{2\xi^2}{(1+\xi^2)^2\hskip-0.1cm}\hskip0.1cm=-\frac2{(1+\xi^2)^2\hskip-0.1cm}\hskip0.1cm+\frac1{1+\xi^2\hskip-0.1cm}$\hskip0.2cm holds, then, for all $m\in\N$,  the last formula  is equivalent to the following
\be\label{AUX-I}\ba{c}\displ\sfrac{1+\tan^2\!\alpha \frac\pi2\hskip-0.1cm} {\tan\alpha \frac\pi2} \!\!\!\displ\intl{J(\alpha ,m)}\!\!\!\sfrac{e_m}{1+\rho_m^2}\dot\rho_md\tau+2\sfrac{1+\tan^2\!\alpha \frac\pi2\hskip-0.1cm} {\tan\alpha \frac\pi2} \!\!\!\intl{J(\alpha ,m)}\!\!\!\sfrac{\rho_m^2}{1+\rho_m^2}d\tau-2\sfrac{1+\tan^2\!\alpha \frac\pi2\hskip-0.1cm} {\tan\alpha \frac\pi2}\!\! \!\intl{J(\alpha ,m)}\!\!\!\sfrac{e_m}{(1+\rho_m^2)^2\hskip-0.1cm}\hskip0.1cm\dot\rho_md\tau\\\displ\hskip6cm  = 2\!\!\intl{J(\alpha ,m)}\!\!\!\!\rho_md\tau\,.\ea\ee
Since  the right-hand side is bounded uniformly in $m\in\N$, for all $\alpha$,  we get 
\be\label{SLM-III}2\,\overline{\lim_m}\intl{J(\alpha ,m)}\!\!\!\!\rho_md\tau=A(\alpha )+B(\alpha )\,,\ee where we set$$\displ A(\alpha ):=\sfrac{1+\tan^2\!\alpha \frac\pi2\hskip-0.1cm} {\tan\alpha \frac\pi2} \lim_m \!\!\!\displ\intl{J(\alpha ,m)}\!\!\!\sfrac{e_m}{1+\rho_m^2}\dot\rho_md\tau\,,\;$$ and \be\label{PSLM} B(\alpha ):=2\,\overline{\lim_m}\Bigg[\sfrac{1+\tan^2\!\alpha \frac\pi2\hskip-0.1cm} {\tan\alpha \frac\pi2} \!\!\!\intl{J(\alpha ,m)}\!\!\!\sfrac{\rho_m^2}{1+\rho_m^2}d\tau- \sfrac{1+\tan^2\!\alpha \frac\pi2\hskip-0.1cm} {\tan\alpha \frac\pi2}\!\! \!\intl{J(\alpha ,m)}\!\!\!\sfrac{e_m}{(1+\rho_m^2)^2\hskip-0.1cm}\hskip0.1cm\dot\rho_md\tau\Bigg]\,.\ee The terms $A(\alpha)$ and $B(\alpha)$ are both well posed. For $A(\alpha)$, via \rf{SLM-II},  we have the existence of the limit on $m$. For $B(\alpha)$     
   both the terms on the right-hand side of \rf{PSLM} are bounded with a constant independent of  $m\in\N$ (the former being $J(\alpha,m)\subset(s,t)$, the latter via \rf{DG-III}). We recall the following elementary limit property:
$$\lim_{\alpha \to1^-}{(1-\alpha )\tan\alpha \sfrac\pi2}=\sfrac2\pi\,.$$
Hence, we get
\be\label{SLM-V}\lim_{\alpha \to1^-}A(\alpha )=  \sfrac2\pi\lim_{\alpha \to1^-}\sfrac1{1-\alpha }\lim_m\intl{J(\alpha ,m)}\sfrac{e_m}{1+\rho_m^2}\dot \rho_md\tau\,. \ee This limit is well posed by  virtue of \rf{LS-I}. Moreover, the lower bound of $\rho_m$, implicit in  \rf{LS-IO}$_2$, and estimate \rf{DG-III}     allow us to deduce that
$$\sfrac{1+\tan^2\!\alpha \frac\pi2\hskip-0.1cm} {\tan\alpha \frac\pi2}\,\,\overline{\lim_m}\, \Big|\!\!\! \!\intl{J(\alpha ,m)}\!\!\!\sfrac{e_m}{(1+\rho_m^2)^2\hskip-0.1cm}\hskip0.1cm\dot\rho_md\tau\Big|\leq \sfrac1{\tan\alpha \frac\pi2}\,\,\overline{\lim_m} \,\Big|\! \!\intl{J(\alpha ,m)}\!\!\!\sfrac{e_m}{1+\rho_m^2\hskip-0.1cm}\hskip0.1cm\dot\rho_md\tau\Big| \leq\sfrac {1+c\dm v_0\dm_2^2}{\tan\alpha \frac\pi2}\,. $$
Hence, one deduces that $$\lim_{\alpha \to1^-}\sfrac{1+\tan^2\!\alpha \frac\pi2\hskip-0.1cm} {\tan\alpha \frac\pi2}\,\,\overline{\lim_m}\,\Big|\!\!\!\!\intl{J(\alpha ,m)}\!\!\!\sfrac {e_m}{(1+\rho_m^2)^2}\dot \rho_md\tau\Big|=0\,.$$
These last and   the limit property  \rf{CRGFIC-I} lead to \be\label{SLM-XX}\lim_{\alpha \to1^-} B(\alpha )=0\,.\ee  
The limits achieved in    \rf{SLM-V} and \rf{SLM-XX}, via \rf{SLM-III}, ensure \rf{SLM-VI} for the upper limit. In analogous way one deduces    the lower limit property in \rf{SLM-VI}.\ep
\begin{rem}{\rm Under the assumptions of Lemma\,\ref{I-LS},  for all $\alpha\in(\alpha_0,1)$, Lemma\,\ref{CDD}  leads to the existence of the upper limit $L(\alpha)$ and lower limit $l(\alpha)$ with respect to $m\in\N$. However, letting $\alpha\to1^-$, their limits are equal, see \rf{SLM-VI}.\par Thus, we  detect two different subsequences indexed by $m^{\null_{''}}$ and by $m^{\null_{'}}$, extracted by the sequence in $m$, one achieving the upper limit $L(\alpha)$ and the other achieving lower limit $l(\alpha)$, respectively. \par In the following, we consider indifferently the two subsequences, and,   abusing in the use of the notation, we use for both the subsequences the same index $m$. Sure that no confusion arises.}\end{rem}
\begin{lemma}\label{MJAM}{\sl Under the  assumptions of Lemma\,\ref{I-LS}, for the subsequences, related to the upper limit and lower limit, respectively,  the following {limit properties} hold:\be\label{MJAM-I}\lim_{\alpha\to1^-}\,(1-\alpha)^{-1}\,{\ov{\lim_{m''}}}\,|J(\alpha,m'')| =\lim_{\alpha\to1^-}\,(1-\alpha)^{-1}\,{\ov{\lim_{m'}}}\,|J(\alpha,m')| =0\,.\ee}\end{lemma}\bp 
For all $h\in \N(\alpha,m)$, we evaluate the energy relation \rf{ER} on the interval $(s_h,t_h)$: 
\be\label{P-V}\sfrac12\sfrac d{dt}e_m(t)+\rho_m(t)=0\,,\quad t\in(s_h,t_h)\,,\,\Longrightarrow\; e_m(t_h)+2\intll{s_h}{t_h}\rho_md\tau=e_m(s_h)\,.\ee
After multiplying this last by $\arctan \rho_m$, we integrate by parts on $(s_h,t_h)$. In virtue of      the values of $\rho_m$   in the endpoints of intervals $(s_h,t_h)$,  stated  in \rf{LS-IO}$_3$, we get $$\alpha\sfrac\pi2\sfrac12\big[e_m(t_h)-e_m(s_h)\big]+\intll{s_h}{t_h}\arctan\rho_m \hskip0.05cm\rho_m d\tau=\sfrac12\intll{s_h}{t_h}\sfrac{e_m\dot\rho_m}{1+\rho^2_m}d\tau\,,$$ that, via \rf{P-V}, is equivalent to
\be\label{P-VI}\intll{s_h}{t_h}\big[\arctan\rho_m -\alpha\sfrac\pi2\big]\rho_m d\tau=\sfrac12\intll{s_h}{t_h}\sfrac{e_m\dot\rho_m}{1+\rho^2_m}d\tau\,,\ee
that has the left-hand side non-negative, for all $h\in \N(\alpha,m)$ and $m\in\N$. We sum \rf{P-VI} on index $h\in\N(\alpha,m)$. Hence, the following holds:
$$\intl{J(\alpha,m)}\big[\arctan\,\rho_m- \alpha\sfrac\pi2 \big]\rho_md\tau=\sfrac12\intl{J(\alpha,m)}\sfrac{e_m\dot\rho_m}{1+\rho^2_m}d\tau\,.$$
We consider the following decomposition:
$$\intl{J(\alpha,m)}\big[\sfrac\pi2-\arctan\,\rho_m\big]\rho_md\tau-(1-\alpha)\sfrac\pi2\intl{J(\alpha,m)}\rho_md\tau=-\sfrac12\intl{J(\alpha,m)}\sfrac{e_m\dot\rho_m}{1+\rho^2_m}d\tau\,.$$ By virtue of \rf{SLM-II}, the  right-hand side admits limit on $m$, and the same limit property holds for any subsequence. Hence,   we get
$${\lim_m}\!\!\!\!\intl{J(\alpha,m)}\!\!\!\!\big[\sfrac\pi2-\arctan\,\rho_m\big]\rho_md\tau=  (1-\alpha)\sfrac\pi2\, {\lim_m}\!\!\! \intl{J(\alpha,m)}\!\!\!\rho_md\tau-\sfrac12\lim_m\!\!\! \intl{J(\alpha,m)}\!\!\!\sfrac{e_m\dot\rho_m}{1+\rho^2_m}d\tau\,.$$
That we write equivalently as
\be\label{LELM}\sfrac1{1-\alpha}\,{\lim_m}\!\!\!\!\intl{J(\alpha,m)}\!\!\!\!\big[\sfrac\pi2-\arctan\,\rho_m\big]\rho_md\tau=  \sfrac\pi2\,{\lim_m}\!\!\!\! \intl{J(\alpha,m)}\!\!\!\!\rho_md\tau-\sfrac12\sfrac1{1-\alpha}\lim_m \!\!\!\!\intl{J(\alpha,m)}\!\!\!\!\sfrac{e_m\dot\rho_m}{1+\rho^2_m}d\tau\,.\ee
Since both the terms on the right-hand side admit limit on $\alpha\to1^-$ and,  by virtue of \rf{SLM-VI}, are equal, we arrive at
$$\lim_{\alpha\to1^-}\sfrac1{1-\alpha}\,{\lim_m}\!\!\!\!\intl{J(\alpha,m)}\!\!\!\!\big[\sfrac\pi2-\arctan\,\rho_m\big]\rho_md\tau= \sfrac\pi2\lim_{\alpha\to1^-}{\lim_m}\!\!\! \!\intl{J(\alpha,m)}\!\!\!\!\rho_md\tau-\sfrac12\lim_{\alpha\to1^-}  \sfrac1{1-\alpha}  \lim_m \!\!\!\!\intl{J(\alpha,m)}\!\!\!\!\sfrac{e_m\dot\rho_m}{1+\rho^2_m}d\tau =0\,.$$
Since for $x>0$ the function $\big[\frac\pi2-\arctan x\big]x$ is positive and concave   with limit $1$ at $\infty$, via \rf{LELM}, for some positive constant $c(\alpha_1)$, we  get
$$0=\lim_{\alpha\to1^-}\sfrac1{1-\alpha}\,{\lim_m}\!\!\!\!\intl{J(\alpha,m)}\!\!\!\!\big[\sfrac\pi2-\arctan\,\rho_m\big]\rho_md\tau\geq c\lim_{\alpha\to1^-}(1-\alpha)^{-1}\,{\ov{\lim_{m}}}{\,|J(\alpha,m)|} \,,$$ which proves the lemma.  \ep
\begin{rem}\label{MmLl}{\rm In the next lemma we are going to state  formula \rf{UP-O}$_1$ with max-limit on the right-hand side. By the same arguments formula \rf{UP-O}$_2$ holds with the min-limit on the right-hand side. For two different reasons, the nature of the limits is not interesting for our aims.  The former is due to the fact that in any case the subsequent limit on $\alpha$ furnishes the same value. The latter is due to the fact  that in the sequel (see Lemma\,\ref{NUP})  we look for a family on one-parameter $\alpha$, say $\texttt{n}''(\alpha)$ and $\texttt{n}'(\alpha)$, in place of the possible $\texttt{n}(\alpha,m'')$ and $\texttt{n}(\alpha,m')$, respectively. So that the nature of the limit in \rf{UP-O} becomes inessential.}\end{rem}  
\begin{lemma}\label{CJAM}{\sl  Under the  assumptions of Lemma\,\ref{I-LS},  the limit value \rf{SLM-VI}, initially stated by considering the  family   $\{J(\alpha,m)\}$ for $\alpha\in(\alpha_1,1)$ and $m\in\N$, holds by considering   a subfamily $\{\widetilde J(\alpha,m^{\null_{''}})\}$  \big(resp. $\{\widetilde J(\alpha,m^{\null_{'}})\}$\big) , for $\alpha\in(\alpha_0',1)$, $\alpha_0'>\alpha_1$,, where $\widetilde J(\alpha,m'')$ \big(resp. $\widetilde J(\alpha,m')$\big) is a set of  pairwise disjoint open intervals with \mbox{\rm\texttt{card}}$(\widetilde J(\alpha,m^{\null_{''}}))\!=$\mbox{\rm\texttt{n}}$(\alpha,m^{\null_{''}})$ $< \lfloor 2c\tan\alpha\frac\pi2\rfloor$  for all $m^{\null_{''}}$ \big(resp. \mbox{\rm\texttt{card}}$(\widetilde J(\alpha,m^{\null_{'}}))\!=\mbox{\rm\texttt{n}}(\alpha,m^{\null_{'}})$ $< \lfloor 2c\tan\alpha\frac\pi2\rfloor$, for all $m^{\null_{'}}$\big) with $t_k-s_k>(\sqrt{2c}\tan{\alpha\frac\pi2})^{-2}$, for all $k=0,\cdots,\mbox{\rm\texttt{n}}(\alpha,m^{\null_{''}})$, \big(resp. $k=0,\cdots,\mbox{\rm\texttt{n}}(\alpha,m^{\null_{'}})$\big), where $c$ is a  constant    independent of $m^{\null_{''}}$ \big(resp. $m^{\null_{'}}\big),v_0$ and $\alpha$. That is, we get
\be\label{UP-O}\ba{l}\displ \sfrac2\pi\lim_{\alpha \to1^-}\!\sfrac1{1-\alpha }\lim_m\!\!\intl{J(\alpha ,m)}\!\!\!\sfrac{e_m}{1\!+\!\rho_m^2\hskip-0.1cm}\hskip0.1cm\dot \rho_{m}d\tau =2\!\lim_{\alpha \to1^-}\!{\lim_{\;\,m^{\null_{''}}}}\!\!\intl{\widetilde J(\alpha ,m^{\null_{''}})}\!\!\!\rho_{m''}d\tau=2\lim_{\alpha \to1^-}\!{\lim_{\;m^{\null_{''}}}}\,\mbox{$\underset{k=1}{\overset{\mbox{\,\scriptsize\rm\texttt{n}}(\alpha,m^{\null_{''}})}\sum}$}\!\intll{s_k}{t_k}\!\rho_{m''}d\tau\,,\VS\sfrac2\pi\lim_{\alpha \to1^-}\!\sfrac1{1-\alpha }\lim_m\!\!\intl{J(\alpha ,m)}\!\!\!\sfrac{e_m}{1\!+\!\rho_m^2\hskip-0.1cm}\hskip0.1cm\dot \rho_{m}d\tau=2\lim_{\alpha \to1^-}{\lim_{\;m^{\null_{'}}}}\intl{\widetilde J(\alpha ,m^{\null_{'}})}\rho_{m'}d\tau\displ =2\lim_{\alpha \to1^-}\!{\lim_{\;m^{\null_{'}}}}\mbox{$\underset{k=1}{\overset{\mbox{\,\scriptsize\rm\texttt{n}}(\alpha,m^{\null_{'}})}\sum}$}\!\intll{s_k}{t_k}\!\rho_{m'}d\tau\,.\ea\ee {Moreover}, 
we get \be\label{LC-I}\ba{c}\displ\lim_{\alpha\to1^-}(1-\alpha)^{-1}\,\ov{\lim_{\;m^{\null_{''}}}}\,|\widetilde J(\alpha,m^{\null_{''}})|=0\mbox{\, and }\lim_{\alpha\to1^-}(1-\alpha)\,{\ov{\lim_{\;m^{\null_{''}}}}}\mbox{\,\rm\texttt n}(\alpha,m^{\null_{''}})=0  \,,\VS \lim_{\alpha\to1^-}(1-\alpha)^{-1}\,{\underset{\,m^{\null_{'}}}{\ov\lim}}\,|\widetilde J(\alpha,m^{\null_{'}})|=0\mbox{\, and }\lim_{\alpha\to1^-}(1-\alpha)\,\ov{\lim_{m^{\null_{'}}}}\mbox{\,\rm\texttt n}(\alpha,m^{\null_{'}})=0\,.\ea\ee}\end{lemma}
\bp We consider an  $\alpha_0\in(\alpha_1,1)$  in such a way that \be\label{MVA}(1-\alpha)\tan{\alpha}\sfrac\pi2<\sfrac2\pi\,,\mbox{ for all }\alpha\in(\alpha_0,1)\,.\ee From now on we restrict our arguments to the parameter $\alpha\in(\alpha_0,1)$. We denote by $J(\alpha,m'')$ the subset of $J(\alpha,m)$ related to the index $m\equiv m''$ of the subsequence $\{\rho_{m''}\}$. In the following of the present proof, by the symbol $c$ we denote the absolute constant that appears in estimate \rf{NSM-IV}.  For an arbitrary $\mu>c$, we define \be\label{UPP-XI}\N^{\null_{''}}(\alpha,m^{\null_{''}}):=\big\{k\in\N(\alpha,m^{\null_{''}}):\,  t_k-s_k\geq \sfrac 1{2\mu\tan^2\alpha\frac\pi2}\mbox{\, holds}\mbox{ for }(s_k,t_k)\in J(\alpha,m^{\null_{''}})\big\}\,.\ee
   Considering the set of indexes $\N^{{''}}(\alpha,m^{\null_{''}})$, we trivially  find a lower bound for $|J(\alpha,m'')|$\,:
$$|J(\alpha,m'')|=\hskip-0.1cm\mbox{$\underset{h\in\N(\alpha,m'')}\sum$}\hskip-0.2cm(t_h-s_h)\geq \hskip-0.1cm\mbox{$\underset{k\,\in\,\N^{\null_{''}}\!(\alpha,m^{\null_{''}})}\sum$}\hskip-0.2cm(t_k-s_k)\,.$$
Assume that $\texttt{card}(\N^{\null_{''}}(\alpha,m^{\null_{''}}))\geq\lfloor 2c\tan \alpha\frac\pi2 \rfloor$ for all $\alpha\in(\alpha_0,1)$. Then we also obtain $$|J(\alpha,m'')|\geq \sfrac{\lfloor 2c\tan\alpha\frac\pi2 \rfloor}{2\mu\tan^2\alpha\frac\pi2 }\,.$$ Recalling that in \rf{MVA} the inequality $(1-\alpha)\tan\alpha\frac\pi2 <\frac2\pi$ holds for all $\alpha\in(\alpha_0,1)$, the last estimate leads to \be\label{ASC}\sfrac2\pi(1-\alpha)^{-1}|J(\alpha,m'')|>\sfrac{\lfloor 2c\tan\alpha\frac \pi2 \rfloor}{2\mu\tan\alpha\frac\pi2 }\,.\ee For some $\alpha_0'\in (\alpha_0,1)$,  via \rf{MJAM-I}, estimate \rf{ASC} is an  absurd for all $\alpha\in(\alpha_0',1)$.  This proves that, for some $\alpha_0'\in(\alpha_0,1)$,\be\label{UP-XI}\texttt{card}(\N^{\null_{''}}(\alpha,m^{\null_{''}}))=\texttt{n}^{\null_{''}}(\alpha,m^{\null_{''}})< \lfloor 2c\tan\alpha\frac\pi2 \rfloor\,,\mbox{ for all }  m''\in\N\mbox{ and }\alpha\in(\alpha_0',1) \,.\ee   
Now, we consider $h\in \N(\alpha,m^{\null_{''}})-\N^{\null_{''}}(\alpha,m^{\null_{''}})$. For such indexes $h$ we have $t_h(\alpha,m^{\null_{''}})-s_h(\alpha,m^{\null_{''}})<\frac1{2\mu\tan^2\alpha\frac\pi2}\,.$ Integrating the differential inequality \rf{NSM-IV}, we get
$$\rho_{m^{\null_{''}}}(\tau)\leq \sfrac{\tan\alpha\frac\pi2 }{\big[1-2c(\tau-s_h)\tan^2\!\alpha\frac\pi2  \big]^\frac12\hskip-0.2cm}\hskip0.2cm\,,\mbox{ for all }\tau\in(s_h(\alpha,m^{\null_{''}}),s_h(\alpha,m^{\null_{''}})+\sfrac1{2c\tan^2\!\alpha\frac\pi2 })\,,\mbox{ for all }m^{\null_{''}}\,,$$ where we taken into account that, for all $h\in\N(\alpha,m'')$  and  $m^{\null_{''}}\in\N$, $\rho_{m^{\null_{''}}}(s_h(\alpha,m^{\null_{''}}))=\tan\alpha\frac\pi2 $ holds.
Thus, being $\mu>c$ and  $t_k-s_k<(2\mu\tan^2\alpha\frac\pi2)^{-1}$, we get $\rho_{m^{\null_{''}}}(\tau)\leq \tan\alpha\frac\pi2\big[1-\frac c\mu]^{-\frac12}$ for all $\tau\in(s_h,t_h)$, that allows us the estimate 
$$ \overline{\lim_{\;m^{\null_{''}}}}\mbox{${\underset{h\in\N(\alpha,m^{\null_{''}})-\N''(\alpha,m^{\null_{''}})}\sum}$}\intll{s_h}{t_h}\hskip-0.1cm\rho_{m^{\null_{''}}}d\tau \leq  \sfrac{\mu^\frac12\!\tan{\alpha\!\frac\pi2}}{(\mu-c)^\frac12}\,\overline{\lim_{m''}}\,|J(\alpha,m'')| \,.$$ Hence,  via \rf{MJAM-I}, for some constant $\eta>0$, the following holds:
\be\label{UP}\lim_{\alpha\to1^-}\overline{\lim_{\;m^{\null_{''}}}}\mbox{${\underset{h\in\N(\alpha,m^{\null_{''}})-\N''(\alpha,m^{\null_{''}})}\sum}$}\intll{s_h}{t_h}\hskip-0.1cm\rho_{m^{\null_{''}}}d\tau\leq\eta\!\lim_{\alpha\to1^-}\!\tan{\alpha\!\sfrac\pi2}\,\overline{\lim_{m''}}\,|J(\alpha,m'')|=0\,.\ee
We consider formula \rf{AUX-I} restricted to the sequence indexed in $\{m^{\null_{''}}\}$:  
$$\ba{c}\displ\sfrac{1+\tan^2\!\alpha \frac\pi2\hskip-0.1cm} {\tan\alpha \frac\pi2} \!\!\!\displ\intl{J(\alpha ,m'')}\!\!\!\sfrac{e_{m''}}{1+\rho_{m''}^2}\dot\rho_{m''}d\tau+2\sfrac{1+\tan^2\!\alpha \frac\pi2\hskip-0.1cm} {\tan\alpha \frac\pi2} \!\!\!\intl{J(\alpha ,m'')}\!\!\!\sfrac{\rho_{m''}^2}{1+\rho_{m''}^2}d\tau-2\sfrac{1+\tan^2\!\alpha \frac\pi2\hskip-0.1cm} {\tan\alpha \frac\pi2}\!\! \!\intl{J(\alpha ,m'')}\!\!\!\sfrac{e_{m''}}{(1+\rho_{m''}^2)^2\hskip-0.1cm}\hskip0.1cm\dot\rho_{m''}d\tau\\\displ\hskip6cm  = 2\intl{J(\alpha,m'')}\hskip-0.1cm\rho_{m^{\null_{''}}}d\tau\,.\ea$$
Recalling the definition set in \rf{UPP-XI}, via \rf{UP-XI}, the last is modified in the following way: 
$$\ba{c}\displ\sfrac{1+\tan^2\!\alpha \frac\pi2\hskip-0.1cm} {\tan\alpha \frac\pi2} \!\!\!\displ\intl{J(\alpha ,m'')}\!\!\!\sfrac{e_{m''}}{1+\rho_{m''}^2}\dot\rho_{m''}d\tau+2\sfrac{1+\tan^2\!\alpha \frac\pi2\hskip-0.1cm} {\tan\alpha \frac\pi2} \!\!\!\intl{J(\alpha ,m'')}\!\!\!\sfrac{\rho_{m''}^2}{1+\rho_{m''}^2}d\tau-2\sfrac{1+\tan^2\!\alpha \frac\pi2\hskip-0.1cm} {\tan\alpha \frac\pi2}\!\! \!\intl{J(\alpha ,m'')}\!\!\!\sfrac{e_{m''}}{(1+\rho_{m''}^2)^2\hskip-0.1cm}\hskip0.1cm\dot\rho_{m''}d\tau\\\displ\hskip6cm  = 2\,\mbox{$\underset{k=1}{\overset{\mbox{\scriptsize\rm\texttt{n}}(\alpha,m^{\null_{''}})}\sum}$}\!\intll{s_k}{t_k}\!\rho_{m''}d\tau+2\mbox{${\underset{h\in\N(\alpha,m^{\null_{''}})-\N''(\alpha,m^{\null_{''}})}\sum}$}\intll{s_h}{t_h}\hskip-0.1cm\rho_{m^{\null_{''}}}d\tau\,.\ea$$
Letting $m''\to\infty$, we arrive at
\be\label{SLMA-III}2\,\ov{\lim_{\;m^{\null_{''}}}}\,\mbox{$\underset{k=1}{\overset{\mbox{\scriptsize\rm\texttt{n}}(\alpha,m^{\null_{''}})}\sum}$}\!\intll{s_k}{t_k}\!\rho_{m''}d\tau=A(\alpha )+C(\alpha )\,,\ee where  we set\footnote{\,Since the sequence $\{h(\alpha,m)\}$ admits limit, see \rf{LS-OI}, its value $A(\alpha)$ holds also on any extract.   Instead,   a finite number of addends alters the term $B(\alpha)$ of   formula \rf{SLM-III}. Thus, its expression in \rf{PSLM} is changed. {\it A priori} we need to distinguish it by considering the limit value of a new term that we denote by $C(\alpha)$. This justifies the formula \rf{SLMA-III}.}$$\displ A(\alpha ):=\sfrac{1+\tan^2\!\alpha \frac\pi2\hskip-0.1cm} {\tan\alpha \frac\pi2} \lim_{\;m^{\null_{''}}} \!\!\!\displ\intl{J(\alpha ,m^{\null_{''}})}\!\!\!\sfrac{e_{m''}}{1+\rho_{m''}^2}\dot\rho_{m''}d\tau\,,\;$$ and $$ C(\alpha )\!:=2\,\overline{\lim_{\;m^{\null_{''}}}}\Bigg[\sfrac{1+\tan^2\!\alpha \frac\pi2\hskip-0.1cm} {\tan\alpha \frac\pi2}\! \!\!\!\!\intl{J(\alpha ,m^{\null_{''}})}\!\!\!\!\!\!\sfrac{\rho_{m''}^2}{1+\rho_{m''}^2}d\tau- \sfrac{1+\tan^2\!\alpha \frac\pi2\hskip-0.1cm} {\tan\alpha \frac\pi2}\!\! \!\!\!\intl{J(\alpha ,m^{\null_{''}})}\!\!\!\!\!\!\sfrac{e_{m^{\null_{''}}}}{(1+\rho_{m''}^2)^2\hskip-0.1cm}\hskip0.1cm\dot\rho_{m''}d\tau-%
\!\!\mbox{${\underset{h\in\N(\alpha,m^{\null_{''}})-\N''(\alpha,m^{\null_{''}})}\sum}$}\intll{s_h}{t_h}\hskip-0.1cm\rho_{m^{\null_{''}}}d\tau\Bigg].$$  
The analogous of the limit property \rf{SLM-XX} holds, that is 
$$\lim_{\alpha\to1^-}\,\overline{\lim_{\;m^{\null_{''}}}}\Big|\sfrac{1+\tan^2\!\alpha \frac\pi2\hskip-0.1cm} {\tan\alpha \frac\pi2}\! \!\!\!\!\intl{J(\alpha ,m^{\null_{''}})}\!\!\!\!\!\!\sfrac{\rho_{m''}^2}{1+\rho_{m''}^2}d\tau- \sfrac{1+\tan^2\!\alpha \frac\pi2\hskip-0.1cm} {\tan\alpha \frac\pi2}\!\! \!\!\!\intl{J(\alpha ,m^{\null_{''}})}\!\!\!\!\!\!\sfrac{e_{m^{\null_{''}}}}{(1+\rho_{m''}^2)^2\hskip-0.1cm}\hskip0.1cm\dot\rho_{m''}d\tau\Big|=0\,.$$
Hence, by virtue \rf{UP},  then we also   prove that 
$$\lim_{\alpha\to1^-}C(\alpha)=0\,.$$  
Hence, set \be\label{IJCF}\widetilde J(\alpha,m^{\null_{''}}):=\{(s_h,t_h): t_h-s_h> (\sqrt{2c}\tan{\alpha\frac\pi2})^{-2}\,,\;h=1,\cdots, n''(\alpha,m^{\null_{''}})\}\,,\ee  we arrive at $$\ba{ll}\displ \sfrac2\pi\lim_{\alpha \to1^-}\!\sfrac1{1-\alpha }\lim_{\;m^{\null_{''}}}\!\!\intl{J(\alpha ,m^{\null_{''}})}\hskip-0.4cm\frac{e_{m''}}{1\!+\!\rho_{m^{\null_{''}}}^2\hskip-0.1cm}\hskip0.1cm\dot \rho_{m^{\null_{''}}}d\tau\hskip-0.3cm&\displ =2\!\lim_{\alpha \to1^-}{\lim_{\;m^{\null_{''}}}}\!\!\intl{\widetilde J(\alpha ,m^{\null_{''}})}\hskip-0.4cm\rho_{m^{\null_{''}}}d\tau =2\!\lim_{\alpha \to1^-}{\lim_{\;m^{\null_{''}}}}\,\mbox{$\underset{k=1}{\overset{\mbox{\scriptsize\rm\texttt{n}}(\alpha,m^{\null_{''}})}\sum}$}\!\intll{s_k}{t_k}\!\rho_{m^{\null_{''}}}d\tau\,.\ea$$
Hence,  we arrive at \rf{UP-O}$_1$. \par Being $\widetilde J(\alpha,m'')\subseteq J(\alpha,m'')$, by virtue of \rf{MJAM-I}, the first statement of \rf{LC-I}$_1$ is immediate.       \par Finally, considering again \rf{ASC}, in the light of \rf{MVA} and \rf{UP-XI}, we also get
\be\label{CARD-I}\sfrac2\pi(1-\alpha)^{-1}|J(\alpha,m)|>\sfrac{\lfloor 2c\tan\alpha\frac\pi2 \rfloor}{2\mu\tan\alpha\frac\pi2 }\geq c(\alpha_1)(1-\alpha)\sfrac{\texttt{n}(\alpha,m)}{2\mu }\,,\mbox{ for all }\alpha\in(\alpha_1,1)\,.\ee Letting $m\to\infty$ and then $\alpha\to1^{-}$, via \rf{MJAM-I}, we arrive at the second statement of \rf{LC-I}$_1$.\par In the case of the sequence indexed  in $\{m^{\null_{'}}\}$ the proof of \rf{UP-O}$_2$ and \rf{LC-I}$_2$ are the same.
\ep
\subsection{The  ``special energy equality'' \rf{CTII-II} for the weak solutions of Theorem\,\ref{CTII}}
We start with the following.\begin{lemma}\label{NO}{\sl Under the assumptions of Lemma\,\ref{I-LS} assume that for some $s,t\in \cal T$ the energy equality \rf{PP-II} does not hold for the weak limit $v$. Then for all $\alpha\in (\alpha'',1)$, $\alpha''\geq\alpha'_0$,   the set $\widetilde J(\alpha,\widetilde m),$  defined in \rf{IJCF}, has {\rm\texttt{card}}$\hskip0.05cm\widetilde J(\alpha,\widetilde m)$={\rm\texttt{n}}$(\alpha,\widetilde m)\in[1,\lfloor 2c\tan\alpha\frac\pi2 \rfloor)$ for all  subsequences of indexes $\{\widetilde m\}\subseteq\{m\}$.
}\end{lemma}
\bp Assume that the statement is false. Let  $\nu_k\to1^-$ be  such that for all $k\in\N$ there exists   an extract $\{\widetilde m\}$ enjoying the property \texttt{card}$\hskip0.05cm\widetilde J(\nu_k,\widetilde m)$=\texttt{n}$(\nu_k,\widetilde m)=0$ for all $\widetilde m$ of the extract sequence.  That is, for all $k$, along the subsequence $\{\widetilde m\}$, we have, at least for a $\mu>c$, that $\N''(\nu_k,\widetilde m)=\emptyset$ (see \rf{UPP-XI} for the definition)%, this also means that \texttt{n}$(\alpha,m)=0$ for all $(\alpha,m)\in (\alpha_0'',1)\times\{\widetilde m\}$
. Hence, being \texttt{n}$(\nu_k,\widetilde m)=0$ for all $\widetilde m$  of the extract sequence $\{\widetilde m\}$,  along this sequence $\{\widetilde m\}$ we arrive at  
\rf{SLMA-III} with the left-hand side null, and with \be\label{EIN-I}\displ A(\nu_k ):=\sfrac{1+\tan^2\!\nu_k \frac\pi2\hskip-0.1cm} {\tan \nu_k \frac\pi2} \lim_{\widetilde m} \!\!\!\displ\intl{J(\nu_k ,\widetilde m)}\!\!\!\sfrac{e_{\widetilde m}}{1+\rho_{\widetilde m}^2}\dot\rho_{\widetilde m}d\tau=\sfrac{1+\tan^2\!\nu_k \frac\pi2\hskip-0.1cm} {\tan \nu_k \frac\pi2} \lim_{\widetilde m}h(\nu_k,\widetilde m)\,,\;\ee thanks to \rf{LS-OI} this last holds because $\{\widetilde m\}$ is extracted from $\{m\}$ and $\{h(\alpha,m)\}$ is convergent. Moreover, being $\N''(\nu_k,\widetilde m)=\emptyset$, the function $C(\alpha)$ in \rf{SLMA-III} becomes  \be\label{EIN-II}D(\nu_k):=
2\,\overline{\lim_{\;\widetilde m}}\Bigg[\sfrac{1+\tan^2\!\nu_k \frac\pi2\hskip-0.1cm} {\tan \nu_k \frac\pi2}\! \!\!\!\!\intl{J(\nu_k ,\widetilde m)}\!\!\!\!\!\!\sfrac{\rho_{\widetilde m}^2}{1+\rho_{\widetilde m}^2}d\tau- \sfrac{1+\tan^2\!\nu_k \frac\pi2\hskip-0.1cm} {\tan \nu_k \frac\pi2}\!\! \!\!\!\intl{J(\nu_k ,\widetilde m)}\!\!\!\!\!\!\sfrac{e_{\widetilde m}}{(1+\rho_{\widetilde m}^2)^2\hskip-0.1cm}\hskip0.1cm\dot\rho_{\widetilde m}d\tau-\!\!\mbox{${\underset{h\in\N(\nu_k,\widetilde m) }\sum}$}\intll{s_h}{t_h}\hskip-0.1cm\rho_{\widetilde m}d\tau\Bigg].\ee
Since the limit properties \rf{UP}  and \rf{SLM-XX} hold for the whole sequence $\{m''\}$,   we prove that for $D(\nu_k)$, defined by  \rf{EIN-II}, the following limit holds
\be\label{EIN-III}\lim_{\nu_k\to1^-}D(\nu_k)=0\,.\ee
Hence, for the hypothesis of absurd   we get  relation \rf{SLMA-III} with the left-hand side null, and, via \rf{EIN-III}, for the right-hand    side of \rf{SLMA-III}    we prove that for the term \rf{EIN-I} holds $\displ\lim_{\nu_k\to1^-}A(\nu_k)=0$. So that,   via \rf{LS-I}, we arrive at
$$0=\dm v(s)\dm_2^2-\dm v(t)\dm_2^2-2\intll {s}t\dm \n v\dm_2^2 d\tau\,. $$ this last contradicts the assumption.
\ep
\begin{lemma}\label{NUP}{\sl Under the  assumptions of Lemma\,\ref{I-LS}, if for some $t>s\in \cal T$  the weak solution $v$ does not enjoy the energy equality, then there exists an one parameter family \mbox{\rm$\texttt{n}(\alpha)$} of positive integers such that
the weak solution $v$  satisfies    the following special energy equality:
\be\label{NUP-I}\dm v(t)\dm_2^2+2\intll st\dm\n v(\tau)\dm_2^2d\tau+\lim_{\alpha\to1^-}\mbox{$\underset{h=1}{\overset{\mbox{\scriptsize\rm\texttt{n}}(\alpha)}\sum}$}\Big[\dm v(s_h(\alpha))\dm_2^2-\dm v(t_h(\alpha))\dm_2^2\Big]=\dm v(s)\dm_2^2\,,\ee where {\rm\texttt{n}}$(\alpha)\leq \lfloor 2c\tan\alpha\frac\pi2\rfloor$  $c$  
is a constant  independent of $v_0,\,s,\,t\in\cal T$ and of $\alpha$. Moreover, set $\widetilde J'(\alpha):=\{(s_h(\alpha),t_h(\alpha))\mbox{ for }h=1,\cdots,\mbox{\rm\texttt{n}}(\alpha)\}$, then one gets \be\label{NUP-II}\displ\lim_{\alpha\to1^-}(1-\alpha)^{-1}|\widetilde J'(\alpha)|=0\mbox{\; and \,}\lim_{\alpha\to1^-}(1-\alpha)\mbox{\,\rm\texttt n}(\alpha)=0\,.\ee}\end{lemma}
\bp We work on the sequence $\{v^{m''}\}$ of Lemma\,\ref{CJAM}, the same argument works considering the sequence $\{v^{m'}\}$. Moreover, for the sake of the simplicity, since   abusing in the notation there is no confusion, we denote $\{v^{m''}\}$ by $\{v^m\}$.  The strategy is to construct an extract from $\{v^m\}$ which in the limit leads to \rf{NUP-I}.\par We are going to prove  that
for all $\alpha\in(\alpha_0',1)$ there exists an extract $\{\widetilde m\}\subseteq\{m''\}$ such that  $$\texttt{card}(J'(\alpha,\widetilde m))=\mbox{\rm\texttt{n}}(\alpha)\mbox{ for all }\widetilde m,$$  with $J'(\alpha,m)\subseteq \widetilde J(\alpha,m)$, where $\widetilde J(\alpha,m)$ is the family of subsets  stated in Lemma\,\ref{CJAM}  in order to deduce \rf{UP-O}.\par By virtue of Lemma\,\ref{CJAM},  the  cardinality of the set  $\widetilde J(\alpha,m)$ is the sequence   of nonnegative  integers  $\{\texttt{n}(\alpha,m)\}$,  which is bounded by $\lfloor 2c\tan\alpha\frac\pi2\rfloor$ for all $\alpha\in(\alpha'_0,1)$. Then, for all $\alpha\in(\alpha_0',1)$, there exists $\texttt{n}(\alpha)\leq\lfloor2c\tan\alpha\frac\pi2\rfloor$ as  $\min$-limit with respect to $m$. We denote by $\{\widetilde m\}\subseteq\{m\}$ the sequence that realizes the min-limit. Since $\{\texttt{n}(\alpha,m)\}$ is a sequence   of integers variable in the finite range of integers $\big\{0,\cdots, \lfloor 2c\tan\alpha\frac\pi2\rfloor\big\}$,   we also realize that the extract  $\texttt{n}(\alpha,\widetilde m)=\texttt{n}(\alpha)>0$ for all $\widetilde m$. We set \texttt{n}$(\alpha)>0$ since we denied the energy equality, that is, by virtue of Lemma\,\ref{NO},   we denied, for all $\alpha\in(\alpha'',1)$, $\alpha''\geq\alpha_0'$, the existence of any subsequence for which    \texttt{n}$(\alpha,m)=0$.\par  Now, we work on the subsequence $\widetilde m$ which realizes the $\min$-limit $\texttt{n}(\alpha)$.  Moreover, we set $J'(\alpha,\widetilde m)$ the subset of $\widetilde J(\alpha,\cdot)$, whose cardinality is $\texttt{n}(\alpha)$ for all $\widetilde m$.  \par 
Thus, 
starting from our sequence of index $\widetilde m$, we are going to construct   $2\texttt{n}(\alpha)$ subsequences, one for each endpoint of the intervals, with the constrain that  each  subsequence  is extracted from the previous one.\par  The first subsequence is  the one related to  $\{s_1(\alpha,\widetilde m)\}$ that admits an extract convergent to some $s_1(\alpha)$,  and the subsequence $\{t_1(\alpha,\widetilde{\widetilde m})\}$, with indexes $\{\widetilde{\widetilde m}\}$  extracted from the last one, admits a limit $t_1(\alpha)$ and going on until to $h=\texttt{n}(\alpha)$\,.  With abuse of the notation, since there is no confusion, we again denote the last extract by $\{\widetilde m\}$. \par Of course, being along any extract $s_h(\alpha,\widetilde m)< t_h(\alpha,\widetilde m)< s_{h+1}(\alpha,\widetilde m)$ for all $h=1,\cdots,\texttt{n}(\alpha)-1$, we also get $s_h(\alpha)\leq t_h(\alpha)\leq s_{h+1}(\alpha)$. So that the last extract has a sequence of indexes in $\widetilde m$ for which, for each $h=1,\cdots,\texttt{n}(\alpha)$, we get $\displ\lim_{\widetilde m}s_h(\alpha,\widetilde m)=s_h(\alpha)$ and $\displ\lim_{\widetilde m}t_h(\alpha,\widetilde m)=t_h(\alpha)$.\par  We call the generic endpoint $z_k(\alpha,\widetilde m)$, $k=1,\dots,2\texttt{n}(\alpha)$. For the $2\texttt{n}(\alpha)$  endpoints $z_k(\alpha,\widetilde m)$  we determine  subsequences of the approximating $\{v^{\widetilde m}\}$ with the same procedure used to establish the limits $s_h(\alpha)$ and $t_h(\alpha)$ for $h=1,\dots,\texttt{n}(\alpha)$. \par To achieve this goal, we distinguish the case of $\OO$ bounded domain from the case of $\OO$ unbounded domain.\par {\it The case of $\OO$ bounded} -- For $h=1,\cdots,\texttt{n}(\alpha)$, recalling the values in $J^{1,2}$-norm of $v^{\widetilde m}(s_h(\alpha,\widetilde m))$ and of $v^{\widetilde m}(t_h(\alpha,\widetilde m))$    stated in \rf{LS-IO}$_3$,  we have \be\label{UP-XII}\{v^{\widetilde m}(z_1(\alpha,\widetilde m))\},\mbox{ with }\dm v^{\widetilde m}(z_1(\alpha,\widetilde m))\dm_2\leq \dm v_0\dm_2\mbox{ and }\dm \n v^{\widetilde m}(z_1(\alpha,\widetilde m))\dm_2=\tan\ssfrac\pi2\alpha\,.\ee Hence, this sequence is contained in a ball of $J^{1,2}(\OO)$, so that, by virtue of the Rellich-Kondrachov  theorem, admits an extract strongly convergent in $L^2(\OO)$ and weakly in $J^{1,2}(\OO)$, whose limits admit the same bounds given in \rf{UP-XII}. We evaluate the Relich-Kondrachov  extract with indexes $\{\widetilde {\widetilde m}\}$ in $z_2(\alpha,\widetilde {\widetilde m})$, that is 
$$\{v^{\widetilde {\widetilde m}}(z_2(\alpha,\widetilde {\widetilde m} ))\},\mbox{ with }\dm v^{\widetilde {\widetilde m}}(z_2(\alpha,\widetilde {\widetilde m}))\dm_2\leq \dm v_0\dm_2\mbox{ and }\dm \n v^{\widetilde {\widetilde m}}(z_2(\alpha,\widetilde {\widetilde m}))\dm_2=\tan\alpha\ssfrac\pi2 \,.$$  This extract is   contained in the same ball of $J^{1,2}(\OO)$, so that admits an extract strongly convergent in $L^2(\OO)$ and weakly in $J^{1,2}(\OO)$. We iterate the procedure until to $k=2\texttt{n}(\alpha)$. With abuse of the notation, since there is no confusion, we again denote the last extract by $\{\widetilde m\}$. In this way, by means of the last subsequence of $\{v^{m''}\}$, we state an extract  $\{v^{\widetilde m}(z_k(\alpha,\widetilde m))\}$ that,  for all $k=1,\cdots,2\texttt{n}(\alpha)$, is  convergent in $L^2(\OO)$.\par For $k=1,\cdots,2\texttt{n}(\alpha)$, we denote by $w_k(\alpha,x)$   the limit in $L^2(\OO)$.  Now, our task is to prove that such limits coincide with the following values of the weak solution $v$:
\be\label{UP-XIV}\ba{cc}w_1(\alpha,x)=v(s_1(\alpha),x)\,,&w_2(\alpha,x)=v(t_1(\alpha))\,,\VS w_3(\alpha,x)=v(s_2(\alpha),x)\,,&w_4(\alpha,x)=v(t_2(\alpha),x)\,,\VS\vdots&\vdots  \VS w_{2\texttt{n}(\alpha)-3}(\alpha,x)=v(s_{\texttt{n}(\alpha)-1}(\alpha),x)\,,&w_{2\texttt{n}(\alpha)-2}(\alpha,x)=v(t_{\texttt{n}(\alpha)-1}(\alpha),x)\,,\VS w_{2\texttt{n}(\alpha)-1}(\alpha,x)=v(s_{2\texttt{n}(\alpha)}(\alpha),x)\,,&w_{2\texttt{n}(\alpha)}(\alpha,x)=v(t_{\texttt{n}(\alpha)}(\alpha),x)\,.\ea\ee In the previous table is implicit the following application: to $k\mbox{ odd}\to s_h(\alpha)$ ($h$ suitable), and to $k\mbox{ even}\to  t_h(\alpha)$ ($h$ suitable)\,.\par
In order to prove \rf{UP-XIV}, first of all we recall that, by virtue of \rf{NSM-I}$_1$, the original sequence $\{v^m\}$ is such that, for all $\varphi\in J^2(\OO)$ and $T>0$,  \be\label{UP-XV}\ba{l}\{(v^m,\varphi)\}\subset C([0,T))\mbox{ is an uniformly equicontinuous and bounded sequence,}\VS\mbox{whose limit }(v(t),\varphi)\in C([0,T)),\mbox{ where }v\mbox{ is the weak solution}.\ea\ee Of course, the same property holds for any extract. Recalling that relatively to    $\{z_k(\alpha,\widetilde m)\}$, if $k$  is odd, for a suitable index $h$\,, we have the property $s_h(\alpha,\widetilde m)\to s_h(\alpha)$,  then  we get
$$\ba{ll} (w_k(\alpha),\varphi)\hskip-0.2cm&\displ=\lim_m(v^{\widetilde m}(z_k(\alpha,\widetilde m)),\varphi)\VSE=\lim_{\widetilde m}\big[(v^{\widetilde m}(s_h(\alpha,\widetilde m))-v^{\widetilde m}(s_h(\alpha)),\varphi)\big]+\lim_{\widetilde m}(v^{\widetilde m}(s_h(\alpha)),\varphi)\VSE=v(s_h(\alpha),\varphi)\,,\ea$$ where we employ \rf{UP-XV} for the former limit and the weak convergence for latter limit. Analogously, 
recalling that if $k$ related to $\{z_k(\alpha,\widetilde m)\}$ is even, for a suitable index $h$\,, we have the property $t_h(\alpha,\widetilde m)\to t_h(\alpha)$, then   we get
    $$\ba{ll}(w_k(\alpha),\varphi)\hskip-0.2cm&\displ=\lim_{\widetilde m}(v^{\widetilde m}(z_k(\alpha,\widetilde m)),\varphi)\VSE=\lim_{\widetilde m}\big[(v^{\widetilde m}(t_h(\alpha,\widetilde m))-v^{\widetilde m}(t_h(\alpha)),\varphi)\big]+\lim_{\widetilde m}(v^{\widetilde m}(t_h(\alpha)),\varphi)\VSE=v(t_h(\alpha),\varphi)\,.\ea$$ Since in both the relations the test function  $\varphi$ is arbitrary, then 
we have proved \rf{UP-XIV}. 
\par{\it The case of $\OO$ unbounded} -- In this case we cannot employ the Rellich-Kondrakov theorem. So that, we recall the technique on the convergence in $L^p$. For this end, we employ the Rellich-Kondrakov   theorem  on the bounded domain $\OO\cap B_R$, and for $\OO\cap B^c_R$ we employ estimate \rf{LDD},  that is uniform with respect to $\widetilde m$, being $\{\widetilde m\}$ extract from $\{ m\}$, and with respect to $t\in [0,\theta_0]$.\par 
Hence, we estimate in the following way:
$$\hskip-0.1cm\ba{ll}\dm v^{\widetilde m}(z_1(\alpha,\widetilde m))-v^p(z_1(\alpha,p))\dm_2\hskip-0.3cm&\leq \!\dm v^{\widetilde m}(z_1(\alpha,\widetilde m))-v^{\widetilde p}(z_1(\alpha,\widetilde p))\dm_{L^2(\OO\cap B_R)}\VSE\hskip4cm+\dm v^{\widetilde m}(z_1(\alpha,\widetilde m))-v^{\widetilde p}(z_1(\alpha,\widetilde p))\dm_{\null_{L^2(\OO\cap B_R^c)}}\VSE \leq\!\dm v^{\widetilde m}(z_1(\alpha,\widetilde m))-v^{\widetilde p}(z_1(\alpha,\widetilde p))\dm_{\null_{L^2(\OO\cap B_R)}}\!\!+2\big[\dm v_0\dm_{\null_{L^2(|x|>R)}}\!\!+c(t)\psi(R)\big].\ea$$ Recalling that $\dm v_0\dm_{\null_{L^2(|x|>R)}}+\psi(R)=o(1)$, $c(t)$ is bounded, the last estimate ensures the Cauchy condition. Hence, the wanted convergence holds. Then one proceeds  as in the case of $\OO$ bounded. So that, we consider the case of $\OO$ unbounded as achieved.\par Now,  the proof follows  the same way for both  the kinds of domains. \par
We consider formula \rf{UP-O} for the last  subsequence extract   from $\{m''\}$:
$$ \displ\sfrac2\pi\lim_{\alpha \to1^-}\!\sfrac1{1-\alpha }\lim_{\widetilde m}\!\!\intl{J(\alpha ,\widetilde m)}\!\!\!\sfrac{e_{\widetilde m}}{1\!+\!\rho_{\widetilde m}^2\hskip-0.1cm}\hskip0.1cm\dot \rho_{\widetilde m}d\tau \displ=2\!\lim_{\alpha \to1^-}\!{\lim_{\;\,\widetilde m}}\!\!\intl{\widetilde J'(\alpha ,\widetilde m)}\!\!\!\rho_{\widetilde m}d\tau=2\lim_{\alpha \to1^-}\!{\lim_{\widetilde m}}\,\mbox{$\underset{k=1}{\overset{\mbox{\,\scriptsize\rm\texttt{n}}(\alpha)}\sum}$}\!\intll{s_k(\alpha,\widetilde  m)}{t_k(\alpha,\widetilde m)}\!\rho_{\widetilde m}d\tau\,.$$ Employing for the last term the formula of energy \rf{ER}, we get \be\label{UP-XVI}\ba{ll}\displ\sfrac2\pi\lim_{\alpha \to1^-}\!\sfrac1{1-\alpha }\lim_{\widetilde m}\!\!\intl{J(\alpha ,\widetilde m)}\!\!\!\sfrac{e_{\widetilde m}}{1\!+\!\rho_{\widetilde m}^2\hskip-0.1cm}\hskip0.1cm\dot \rho_{\widetilde m}d\tau\hskip-0.2cm& \displ=\lim_{\alpha \to1^-}\!{\lim_{\widetilde m}}\,\mbox{$\underset{k=1}{\overset{\mbox{\,\scriptsize\rm\texttt{n}}(\alpha)}\sum}$}\Big[\dm v^{\widetilde m}(s_k(\alpha,\widetilde m))\dm_2^2-\dm v^{\widetilde m}(t_k(\alpha,\widetilde m))\dm_2^2\Big]\VSE=\lim_{\alpha \to1^-}\,\mbox{$\underset{k=1}{\overset{\mbox{\,\scriptsize\rm\texttt{n}}(\alpha)}\sum}$}\Big[\dm v(s_k(\alpha))\dm_2^2-\dm v(t_k(\alpha))\dm_2^2\Big]\,.\ea\ee
Being $\displ\lim_{\alpha\to1^-}(1-\alpha)
\sfrac{1+\tan^2\!\alpha \frac\pi2\hskip-0.1cm} {\tan\alpha \frac\pi2} =\sfrac2\pi$\,, substituting \rf{UP-XVI} in \rf{LS-I}, we achieve \rf{NUP-I}. \par  We set $\widetilde J'(\alpha):=\{(s_h(\alpha),t_h(\alpha))\,,$ for all indexes $h=1,\cdots,\mbox{\rm\texttt n}(\alpha)\}\,$ such that   $[s_h(\alpha),t_h(\alpha)]$ is not a degenerate interval. The set $J'(\alpha)$ is not empty. Because if we get  $s_h(\alpha)\equiv t_h(\alpha)$ for all indexes $h$, then by virtue of \rf{UP-XIV} we also get that  the gap $\dm v(s_h(\alpha)\dm_2-\dm v(t_h(\alpha)\dm_2=0\,.$ This last would furnish the energy equality on $(s,t)$, which contradicts the assumption.  \par Therefore, we can define an application 
$${\texttt{n}}'\!:\alpha\in\!(\alpha_0',1)\to \N: {\texttt{n}'(\alpha)}=\mbox{number of indexes }h\in\!\{1,\cdots,{\texttt{n}(\alpha)}\}\mbox{ such that } s_h(\alpha)\ne t_h(\alpha)\,.$$\par Since there is no confusion, we denote the application ${\texttt{n}}'(\alpha)$ again by ${\texttt{n}(\alpha)}$.\par Recalling the definition of $\{(s_h(\alpha),t_h(\alpha))\}$ and the one related to $J(\alpha,m)$, we easily get
$$\ba{ll}|\widetilde J'(\alpha)|\hskip-0.2cm&=\Big|\,\mbox{\large$\underset{h=1}{\overset{\texttt{n}(\alpha)}\sum}$}(t_h(\alpha)-s_h(\alpha))\Big|=\Big|\,\mbox{\large$\underset{h=1}{\overset{\texttt{n}(\alpha)}\sum}$}(t_h(\alpha)-t_h(\alpha,\widetilde m)+t_h(\alpha,\widetilde m)-s_h(\alpha,\widetilde m)+s_h(\alpha,\widetilde m)-s_h(\alpha))\Big|\VSE \leq\Big|\,\mbox{\large$\underset{h=1}{\overset{\texttt{n}(\alpha)}\sum}$}(t_h(\alpha)-t_h(\alpha,\widetilde m))\Big|+\Big|\,\mbox{\large$\underset{h=1}{\overset{\texttt{n}(\alpha)}\sum}$}(t_h(\alpha,\widetilde m)-s_h(\alpha,\widetilde m))\Big|+\Big|\,\mbox{\large$\underset{h=1}{\overset{\texttt{n}(\alpha)}\sum}$}(s_h(\alpha,\widetilde m)-s_h(\alpha))\Big|\VSE\leq \Big|\,\mbox{\large$\underset{h=1}{\overset{\texttt{n}(\alpha)}\sum}$}(t_h(\alpha)-t_h(\alpha,\widetilde m))\Big|+ |J(\alpha,\widetilde m)|+\Big|\,\mbox{\large$\underset{h=1}{\overset{\texttt{n}(\alpha)}\sum}$}(s_h(\alpha,\widetilde m)-s_h(\alpha))\Big| \,.\ea$$
Since $\{s_h(\alpha,\widetilde m)\}\to s_h(\alpha)$ and $\{t_h(\alpha,\widetilde m)\}\to t_h(\alpha)$, letting $\widetilde m\to\infty$, we arrive at
$$|\widetilde J(\alpha)|\leq \overline{\lim_{m}}\,|J(\alpha,m)|\,.$$ Multiplying by $(1-\alpha)^{-1}$, by virtue of \rf{MJAM-I}, we achieve the proof of the former claim in  \rf{NUP-II}. In order to prove the latter claim of \rf{NUP-II}, one considers again \rf{CARD-I} that along the sequence $\{\widetilde m\}$  becomes $$\sfrac2\pi\sfrac {|J(\alpha,\widetilde m)|}{1-\alpha}>\sfrac{\lfloor 2c\tan\alpha\frac\pi2 \rfloor}{2\mu\tan\alpha\frac\pi2 }\geq c(\alpha_1)(1-\alpha)\sfrac{\texttt{n}(\alpha,\widetilde m)}{2\mu}=c(\alpha_1)(1-\alpha)\sfrac{\texttt{n}(\alpha)}{2\mu }\,,\mbox{\,for\,all\,}\alpha\!\in\!(\max\{\alpha_1,\alpha'_0\},1).$$ Hence, by letting $\widetilde m\to\infty$ and then  letting $\alpha\to1^-$, via \rf{MJAM-I}, we achieve the latter 
claim of \rf{NUP-II}.
\ep
\subsection{\label{PPTC}Proofs of Proposition\,\ref{PPE}, Theorem\,\ref{CTI} and Corollary\,\ref{CoroC}}
\bp[Proof of Proposition\,\ref{PPE}] The necessary condition is immediate. Actually, given two instants $t>s$ both in $\cal T$, evaluating \rf{PP-II} in $t$ and in $s$, respectively,  then by the  difference one deduces the energy equality on $(s,t)$. For the sufficient condition, if $0\notin \cal T$, we  consider any sequence $\{s_p\}\subset \cal T$ converging to 0. Hence, by the hypothesis \rf{PP-II}  and property \rf{SLO}, letting $s_p\to0$, we arrive at \rf{PP-II}.\ep
\bp[Proof of Theorem\,\ref{CTI}] The proof is a consequence of Lemma\,\ref{EX-T} and Lemma\,\ref{NO} - Lemma\,\ref{NUP}\,.\ep
\bp[Proof of Corollary\,\ref{CoroC}]  The necessary condition is a consequence  of formula \rf{CTII-II} that implies ${\rm\texttt n}(\alpha)=0$, that is  $\widetilde J'(\alpha)=\emptyset$.   The converse is a consequence of Proposition\,\ref{PPE} and of Lemma\,\ref{NUP}.\ep
\begin{rem}\label{FRE}{\rm A possible result consequence of Lemma\,\ref{NUP} is the following. Actually,  we can work on the subsequence in $\{m\}$ that leads to the definition of $\texttt{n}(\alpha)$ in order to prove \rf{EE-N}. Working in this way,  formula  \rf{UP-O} becomes:
$$\ba{ll}\displ \sfrac2\pi\lim_{\alpha \to1^-}\!\sfrac1{1-\alpha }\lim_m\!\!\intl{J(\alpha ,m)}\!\!\!\sfrac{e_m}{1\!+\!\rho_m^2\hskip-0.1cm}\hskip0.1cm\dot \rho_md\tau\hskip-0.3cm&\displ =2\lim_{\alpha \to1^-} \mbox{$\underset{k=1}{\overset{\mbox{\scriptsize\rm\texttt{n}}(\alpha)}\sum}$}\,{\lim_{m}}\intll{s_k(\alpha,m)}{t_k(\alpha,m)}\!\rho_md\tau\,.\ea$$ Hence, we can modify \rf{EE-N}. That is, we get  
\be\label{EE-N-I}\dm v(t)\dm_2^2+2\!\intll st\dm\n v(\tau)\dm_2^2d\tau-\dm v(s)\dm_2^2=-2\lim_{\alpha \to1^-} \mbox{$\underset{k=1}{\overset{\mbox{\scriptsize\rm\texttt{n}}(\alpha)}\sum}$}{\,\lim_{m}}\intll{s_k(\alpha,m)}{t_k(\alpha,m)}\!\rho_{m}d\tau\,,\ee where $\texttt{n}(\alpha)$   and  the extract $\{\rho_m\}$  are  deduced as in the  proof of  \rf{NUP-I}\,.}\end{rem}
\vskip0.5cm 
{\bf Declaration:\vskip0.1cm\par Conflict of interests} The author declares no conflict of interest.

\end{document}